\documentclass[10pt,journal,compsoc]{IEEEtran}
\usepackage[numbers,sort]{natbib}
\usepackage{comment}
\usepackage{adjustbox}
\usepackage[table,xcdraw]{xcolor}

\usepackage{anyfontsize}
\usepackage{url}
\usepackage{booktabs}
\usepackage{enumitem}
\setlist[itemize]{leftmargin=*}

\usepackage[caption=false]{subfig}
\usepackage{amsmath,epsfig,psfrag} 
\usepackage{algorithm}
\usepackage{algpseudocode}
\usepackage{booktabs}
\usepackage[normalem]{ulem}

\usepackage{standalone}
\usepackage{tikz}
\usetikzlibrary{arrows}
\usetikzlibrary{automata}
\usetikzlibrary{shapes}
\usetikzlibrary{backgrounds}
\usetikzlibrary{calc}
\usetikzlibrary{fit}
\usetikzlibrary{matrix}
\usetikzlibrary{petri}
\usetikzlibrary{mindmap}
\usetikzlibrary{positioning}
\usetikzlibrary{shadows}
\usetikzlibrary{decorations}
\usetikzlibrary{decorations.markings}
\usetikzlibrary{scopes}

\usepackage{array}
\newcolumntype{P}[1]{>{\centering\arraybackslash}p{#1}}
\newcommand{\fbf}{\textbf}

\begin{document}

\title{Foureye: Defensive Deception based on Hypergame Theory Against Advanced Persistent Threats}

\author{Zelin Wan, Jin-Hee Cho,~\IEEEmembership{Senior Member, IEEE}, Mu Zhu, Ahmed H. Anwar, Charles Kamhoua,~\IEEEmembership{Senior Member, IEEE}, and Munindar P. Singh,~\IEEEmembership{IEEE Fellow} 
\IEEEcompsocitemizethanks{\IEEEcompsocthanksitem Zelin Wan and Jin-Hee Cho are with the Department of Computer Science, Virginia Tech, Falls Church, VA 22043, USA. Email: \{zelin, jicho\}@vt.edu. Mu Zhu and Munindar P. Singh are with the Department of Computer Science, North Carolina State University, Raleigh, NC 27695, USA. Email: \{mzhu5, mpsingh\}@ncsu.edu. Ahmed H. Anwar and Charles A. Kamhoua are with the US Army Research Laboratory, Adelphi, MD 20783, USA. Email: a.h.anwar@knights.ucf.edu; charles.a.kamhoua.civ@mail.mil.}}

\maketitle
\begin{abstract}
Defensive deception techniques have emerged as a promising proactive defense mechanism to mislead an attacker and thereby achieve attack failure.  However, most game-theoretic defensive deception approaches have assumed that players maintain consistent views under uncertainty. They do not consider players' possible, subjective beliefs formed due to asymmetric information given to them. In this work, we formulate a hypergame between an attacker and a defender where they can interpret the same game differently and accordingly choose their best strategy based on their respective beliefs.  This gives a chance for defensive deception strategies to manipulate an attacker's belief, which is the key to the attacker's decision making.  We consider advanced persistent threat (APT) attacks, which perform multiple attacks in the stages of the cyber kill chain where both the attacker and the defender aim to select optimal strategies based on their beliefs. Through extensive simulation experiments, we demonstrated how effectively the defender can leverage defensive deception techniques while dealing with multi-staged APT attacks in a hypergame in which the imperfect information is reflected based on perceived uncertainty, cost, and expected utilities of both attacker and defender, the system lifetime (i.e., mean time to security failure), and improved false positive rates in detecting attackers.
\end{abstract}


\IEEEpeerreviewmaketitle
\pagestyle{plain}
\thispagestyle{plain}

\section{Introduction}
\label{sec:intro}

The key purpose of a defensive deception technique is to mislead an attacker's view and make it choose a suboptimal or poor action for the attack failure~\cite{sharp2006MilitaryDeception}. When both the attacker and defender are constrained in their resources, strategic interactions can be the key to beat an opponent.  In this sense, non-game-theoretic defense approaches have inherent limitations due to lack of efficient and effective strategic tactics.  Forms of deception techniques have been discussed based on certain classifications, such as {\em hiding the truth} vs.\ {\em providing false information} or {\em passive} vs.\ {\em active} for increasing attackers' ambiguity or confusion~\cite{almeshekah2016cyber, caddell2004deception}. 

Game theory has been substantially used for dynamic decision making under uncertainty, assuming that players have consistent views. However, this assumption fails as players may often subjectively process asymmetric information available to them~\cite{Kovach15}. Hypergame theory~\cite{Bennett77} is a variant of game theory that provides a form of analysis considering each player's subjective belief, misbelief, and perceived uncertainty and accordingly their effect on decision making in choosing a best strategy~\cite{Kovach15}.

This paper leverages hypergame theory to resolve conflicts of views of multiple players as a robust decision-making mechanism under uncertainty where the players may have different beliefs towards the same game. Hypergame theory models players, such as attackers and defenders in cybersecurity to deal with advanced persistent threat (APT) attacks.  We dub this effort {\em Foureye} after the {\em Foureye butterflyfish}, demonstrating deceptive defense in nature~\cite{foureye}.

To be specific, we identify the following nontrivial challenges in obtaining a solution. First of all, it is not trivial to derive realistic game scenarios and develop defensive deception techniques to deal with APT attacks beyond the reconnaissance stage. This aspect has not been explored in the state-of-the-art. Second, quantifying the degree of uncertainty in the views of attackers and defenders is challenging, although they are critical because how each player frames a game significantly affects its strategies to take. Third, given a number of possible choices under dynamic situations, dealing with a large number of solution spaces is not trivial whereas the deployment and maintenance of defensive deception techniques is costly in contested environments. We partly addressed these challenges in our prior work in~\cite{Cho19-hgt}; however, its contribution is very limited in considering a small-scale network and a small set of strategies with a highly simplified probability model developed using Stochastic Petri Network. 

To be specific, this paper has the following {\bf new key contributions}:
\begin{itemize}
\item We modeled an attack-defense game under uncertainty based on hypergame theory where an attacker and a defender have different views of the situation and are uncertain about strategies taken by their opponents. 

\item We reduced a player's action space by using a subgame determined based on a set of strategies available where each subgame is formulated based on each stage of the cyber kill chain (CKC) based on a player's belief under uncertainty. 

\item We considered multiple defense strategies, including defensive deception techniques whose performance can be significantly affected by an attacker's belief and perceived uncertainty, which impacts its choice of a strategy.

\item We modeled an attacker's and a defender's uncertainty towards its opponent (i.e., the defender and the attacker, respectively) based on how long each player has monitored the opponent and its chosen strategy. To the best of our knowledge, prior research on hypergame theory uses a predefined constant probability to represent a player's uncertainty. In this work, we estimated the player's uncertainty based on the dynamic, strategic interactions between an attacker and a defender.

\item We conducted comparative performance analysis with or without a defender using defensive deception (DD) strategies and with or without perfect knowledge available towards actions taken by the opponent. We measured the effectiveness and efficiency of DD techniques in terms of a system's security and performance, such as perceived uncertainty, hypergame expected utility, action cost, mean time to security failure (MTTSF or system lifetime), and improved false positive rate (FPR) of an intrusion detection by the DD strategies taken by the defender.
\end{itemize}
\vspace{-4mm}
\section{Related Work} \label{sec:related_work}

\citet{garg2007} proposed a game-theoretic deception framework in honeynets with imperfect information to find optimal actions of an attacker and a defender and investigated the mixed strategy equilibrium. \citet{carroll2011} used deception in attacker-defender interactions in a signaling game based on perfect Bayesian equilibria and hybrid equilibria. They considered defensive deception techniques, such as honeypots, camouflaged systems, or normal systems. 
\citet{yin2013} considered a Stackelberg attack-defense game where both players make decisions based on their perceived observations and identified an optimal level of deceptive protection using fake resources.  \citet{casey2018CACM} examined how to discover Sybil attacks based on an evolutionary signaling game where a defender can use a fake identity to lure the attacker to facilitate cooperation. \citet{schlenker2018ICAAMS} studied a sophisticated and na\"{i}ve APT attacker in the reconnaissance stage to identify an optimal defensive deception strategy in a zero-sum Stackelberg game by solving a mixed integer linear program. 

Unlike the above works cited~\cite{garg2007, carroll2011, yin2013, casey2018CACM, schlenker2018ICAAMS}, our work used hypergame theory which offers the powerful capability to model uncertainty, different views, and bounded rationality by different players. This way reflects more realistic scenarios between the attacker and defender.  

Hypergame theory has emerged to better reflect real-world scenarios by capturing players' subjective and imperfect belief, aiming to mislead them to adopt uncertain or non-optimized strategies. Although other game theories deal with uncertainty by considering probabilities that a certain event may happen, they assume that all players play the same game~\cite{Tadelis13}. Hypergame theory has been used to solve decision-making problems in military and adversarial environments~\citet{Vane99, Vane05, House10}. Several studies~\cite{Gharesifard10,Gharesifard12} investigated how players' beliefs evolve based on hypergame theory by developing a misbelief function measuring the differences between a player's belief and the ground truth payoff of other players' strategies. \citet{Kanazawa07} studied an individual's belief in an evolutionary hypergame and how this belief can be modelled by interpreter functions. \citet{Sasaki14} discussed the concept of {\em subjective rationalizability} where an agent believes that its action is a best response to the other agent's choices based on its perceived game. 

\citet{Putro00} proposed an adaptive, genetic learning algorithm to derive optimal strategies by players in a hypergame. \citet{ferguson2019SHTSS} studied the placement of decoys based on a hypergame. This work developed a game tree and investigated an optimal move for both an attacker and defender in an adaptive game. \citet{Aljefri17} studied a first level hypergame involving misbeliefs to resolve conflicts for two and then more decision makers.  \citet{bakker2019ICSL} modeled a repeated hypergame in dynamic stochastic setting against APT attacks primarily in cyber-physical systems. 

Unlike the works using hypergame theory above~\cite{Aljefri17, bakker2019ICSL, ferguson2019SHTSS, Gharesifard10, Gharesifard12, House10, Kanazawa07, Putro00, Sasaki14, Vane99, Vane05}, our work considered an APT attacker performing multi-staged attacks where attack-defense interactions are modeled based on repeated hypergames. In addition, we show the effectiveness of defensive deception techniques by increasing the attacker's uncertainty leading to choosing non-optimal actions and increasing the quality of the intrusion detection (i.e., a network-based intrusion detection system, NIDS) through the collection of attack intelligence using defensive deception strategies. 
\vspace{-4mm}
\section{System Model} \label{sec:system-model}

\subsection{Network Model}
\label{subsec:network-model}

This work concerns a software-defined network (SDN)-based Internet-of-Things (IoT) environment characterized by servers and/or IoT devices, such as an SDN-based smart environment~\cite{Boussard15}. The key benefit of using the SDN technology is decoupling the network control plane from the data plane (e.g., packet forwarding) for higher flexibility, robust security/performance, and programmability for a networked system in which an SDN controller can efficiently and effectively manage security and performance mechanisms. We use the SDN controller to involve packet forwarding decisions and to deploy defense mechanisms, such as firewalls or NIDSs.  SDN-enabled switches handle forwarding packets, where they encapsulate packets without exact matching flow rules in flow tables in which the encapsulated packets, `OFPT PACKET IN' packets in OpenFlow (OF) protocol (i.e., a standard communication protocol between SDN-enabled switches and the SDN controller), are provided to the SDN controller handling the flow. 

The nodes in this environment collect data and perform a periodic delivery of those collected data to the servers via multi-hop communications, in which the servers may need to process further to provide queried services. The nodes may be highly heterogeneous in their types and functionalities and spread over different Virtual Local Area Networks (VLANs) of the IoT environment. Each VLAN may have one or more servers and is assigned with a set of nodes based on the common characteristics of their functionalities. We leverage the advanced SDN technology~\cite{Macedo15-sdn-survey} for the effective and efficient management of IoT nodes with the help of an SDN controller.


\subsection{Node Model} \label{subsec:node-model}

A node, including web servers, databases, honeypots, and IoT devices, is characterized by the following set of features:
\begin{itemize}
\item {\em Criticality:} This metric, $\mathrm{c}_i$, indicates how critical node $i$ is in terms of its given role for security and reachability (i.e., influence) in a network to maintain network connectivity, and given by: 
\begin{equation}
\vspace{-1mm}
\mathrm{c}_i = \mathrm{importance}_i \times  \mathrm{reachability}_i,  
\label{eq:node-c_i}
\end{equation}
where $\mathrm{importance}_i$ is given as an integer ranged in $[0, 10]$ during the network deployment phrase. $\mathrm{reachability}_i$ is computed based on the faster betweenness centrality metric~\cite{brandes2001faster} by the SDN controller. 
Note that the algorithmic complexity of the faster betweenness in this work is $O(|V|^2)$ as a given network follows Erd\"{o}s–R\'{e}nyi (ER) network model~\cite{Newman10}. $\mathrm{reachability}_i$ is estimated in the range of $[0, 1]$ as a real number. 
\item {\em Security vulnerability}: A node's vulnerabilities to various types of attacks are considered based on three types of vulnerabilities: (1) vulnerabilities associated with software installed in each node, denoted by $\mathrm{sv}_i$; (2) vulnerabilities associated with encryption keys (e.g., secret or private keys), denoted by $\mathrm{ev}_i$. As a longer-term key exposes higher security vulnerability, the attacker can exploit encryption vulnerability over time with $\hat{\mathrm{ev}}_i = \mathrm{ev}_i \cdot e^{-1/\mathrm{T_{rekey}}}$ and $\mathrm{T_{rekey}}$ is the time elapsed since the attacker has investigated a given  key; and (3) an unknown vulnerability, denoted by $\mathrm{uv}_i$, representing the average unknown vulnerability. We assume that all the vulnerabilities are computed based on the Common Vulnerability Scoring System (CVSS)~\cite{CVSS2018} with the severity value in $[0, 10]$ as an integer.  We measure the average vulnerability associated with node $i$ being vulnerable by: 
\begin{equation}
\mathrm{vulnerability}_i = \frac{\sum_{v_j \in V_i} v_j}{|V_i|},
\label{eq:vulnerability}
\end{equation}
where $V_i$ is a set of vulnerabilities associated with node $i$ (e.g., $\{\mathrm{sv}_0, \mathrm{sv}_1, \mathrm{sv}_2, \mathrm{ev}_0, \mathrm{ev}_1, \mathrm{ev}_2, \mathrm{uv}_0\}$), $v_j$ refers to one of vulnerabilities, associated with node $i$ where $v_j$ is measured based on $[0, 10]$ following the CVSS. We denote $P_i^v = \mathrm{vulnerability}_i/10$ as a normalized vulnerability probability.  $P_i^v$ is used as the probability to exploit (i.e., compromise) node $i$ by an attacker.

\item {\em Mobility}: We model the mobility rate of node $i$ by considering a rewiring probability $P^r_i$ only for IoT devices where node $i$ can be connected with a new IoT node with $P^r_i$. For rewiring connections, node $i$ will select one of its neighbors with $P^r_i$ to disconnect and then select a new node to be connected to maintain a same number of neighbors (nodes being directly connected).  
\end{itemize}
\begin{table}[t]
\centering
\scriptsize  
\caption{\sc Example Node Characteristics.}
\label{tab:node-characteristics}
\vspace{-4mm}
\begin{tabular}{ P{1.5cm} P{1.5cm} P{1.5cm} P{1.5cm} }
\toprule
& Importance & Software Vul. & Encryption Vul. \\
\midrule
Web servers & $[8, 10]$ & $[3, 7]$ & $[1, 3]$ \\
Datbases & $[8, 10]$ & $[3, 7]$ & $[1, 3]$ \\
Honeypots & 0 & $[7, 10]$ & $[9, 10]$ \\
IoT devices & $[1, 5]$ & $[1, 5]$ & $[5, 10]$ \\
\bottomrule
\end{tabular}
\vspace{-5mm}
\end{table}
Table~\ref{tab:node-characteristics} shows an example set of node characteristics showing the ranges of each node type's attributes and the shown values used as default settings for our experiments in Section~\ref{sec:results-analysis}. We select each attribute value at random based on uniform distribution in a given range.  Notice that we consider zero importance for honeypots, implying no performance degradation and security damage upon its compromise. In addition, we put a fairly high range of the number of vulnerabilities in the honeypots in order to lure attackers with high attack utility. Since a legitimate user can be compromised by the attacker, $\mathrm{cp}$ refers to the status of a node's compromise (i.e., $\mathrm{cp}_i =1$ for compromise; 0 otherwise).  We summarize node $i$'s profile as:
\begin{equation}
n_i = [\mathrm{c}_i, \mathrm{cp}_i, \mathrm{ev}_i, \mathbf{V}_i, P_i^v, P^r_i].
\label{eq:node-profile}    
\end{equation}
Recall that $\mathrm{c}_i$, $\mathrm{cp}_i$, $\mathrm{ev}_i$, $\mathbf{V}_i$, $P_i^v$, and $P^r_i$ are node $i$'s criticality in $[0, 1]$, the status of being compromised (=1) or not(=0), evicted (=1) or not (=0), vulnerability vector in software, encryption, and unknowns, the probability of the overall vulnerability, and rewiring probability for mobility in $[0, 1]$. 
\vspace{-2mm}
\subsection{Assumptions} \label{subsec:assumptions}

We assume that the SDN controller and control channel are trusted and considering their security vulnerabilities is beyond the scope of this work.  Since each SDN controller should be well informed of basic network information under its control and other SDN controllers' control, each SDN controller periodically updates the network topology and software vulnerabilities of nodes under its control to other SDN controllers. Via this process, each SDN controller can periodically check an overall system security state and take actions accordingly.

We also assume that a network-based IDS (NIDS) is deployed in the SDN controller and is characterized by the probabilities of false positives ($P_{fp}$) and false negatives ($P_{fn}$). The NIDS runs throughout the system lifetime. 
The NIDS's $P_{fp}$ and $P_{fn}$ will be dynamically updated as it receives more attack intelligence from the defensive deception techniques used in this work. We assume that the collected signatures from the deception-based monitoring mechanisms can decrease $P_{fn}$ due to an increased volume of additional signatures.  We simply use Beta distribution to derive $\mbox{Beta}(P_{fn}; \alpha, \beta)$ where $\alpha$ refers to false negatives (FN) and $\beta$ is true positives (TP) with $P_{fn} = FN/(TP+FN)$. Similarly, as more attack intelligence is forwarded to NIDS via defensive deception-based monitoring, $\beta$ (TP) increments by 1 per monitoring interval. Similarly, false positives will be reduced as defensive deception techniques are used where $P_{fp} = FP/(TN+FP)$ and TN increases by 1.

We assume that legitimate users use a secret key for secure group communications among internal, legitimate users while prohibiting outsiders from accessing secured network resources. If an outsider wants to access a target network and become an inside attacker with legitimate credentials, it needs to be authenticated and given the secret key to access the target network. In addition, network resources are accessed according to the privilege of each user. Therefore, to compromise a legitimate node, the attacker should obtain appropriate privileges to access them.
\vspace{-4mm}
\subsection{Attack Model}
\label{subsec:attack-model}
We consider APT attackers performing multi-staged attacks following the cyber kill chain (CKC) for compromising a target node and exfiltrating confidential information to outside~\cite{okhravi2013survey}. We consider the APT attacks as follows. 

{\bf APT Attack Procedure to Achieve Data Exfiltration}: We define an APT attacker's goal in that the attacker has reached and compromised a target node and successfully exfiltrated its confidential data. We assume that nodes with a higher importance (i.e., having more important, credential information) are more likely to be targeted. 

To reach a target node, the attacker needs to compromise other intermediate nodes along the way. We often call the path to the target node `an attack path.' In reality, the attacker may not have an exact, complete view on network topology.  We assume that the attacker only knows its adjacent nodes (i.e., nodes being directly connected) and needs to choose which node to compromise next. The attacker will consider how easily given adjacent node $i$ can be exploited according to an attack cost metric, $ac_k$, for attack strategy $k$. Moreover, if the attacker finds already compromised, adjacent nodes, it can leverage it and has no need to put additional effort to compromise it. We call this `the value of an intermediate node $i$ in an attack path,' denoted by $APV (i, k)$, where $k$ refers to attack strategy ID. Highest $APV (i, k)$ will be added to the attacker's attack path to the target node. $APV (i, k)$ is given by: 
\begin{equation}
APV (i, k) = 
\begin{cases}
(1- \mathrm{\hat{ac}}_k) \cdot P_i^v \; \; \text{if } \mathrm{cp}_i ==0, \\
1 \; \; \text{otherwise.}
\end{cases}
\label{eq:attack-INT-select}
\end{equation}
Here $\mathrm{\hat{ac}}_k = e^{-(1/\mathrm{ac}_k)} \in [0, 1]$ that represents a normalized attack cost where $\mathrm{ac}_k$ is a predefined attack cost ranged in $[0, 3]$ (see `Attack Strategy Attributes' later in this section), and $\mathrm{vulnerability}_i$ is the overall vulnerability in Eq.~\eqref{eq:vulnerability}. Given a node to be compromised next, its vulnerability degree can be computed as $P_{i}^{v}$ ($=\mathrm{vulnerability}_i/10$). If $\mathrm{cp}_i =1 $ (i.e., node $i$ is compromised), the attacker may add it to the attack path at no cost, which gives $APV (i, k) = 1$.  The attacker may need to compromise more than one intermediate nodes before reaching a target node.  

{\bf Attack Strategy Attributes}: An APT attacker can perform multiple attacks through the stages of the CKC. Each attack strategy $k$ can be characterized by: (1) attack cost, $\mathrm{ac}_k$, indicating how much time/effort is needed to launch the attack; and (2) the expected impact (i.e., attack effectiveness) upon attack success, $\mathrm{ai}_k$. $\mathrm{ac}_k$ is a predefined constant as an integer in $[0, 3]$ reflecting no, low, medium, and high cost, respectively. $\mathrm{ai}_k$ is obtained by victim $j$'s criticality, $\mathrm{c}_j$ (see Eq.~\eqref{eq:node-c_i}).  This implies the attack benefit through compromising a set of exploitable  nodes. If there have been multiple nodes being compromised by taking given attack $k$, $\mathrm{ai}_k$ captures the criticalities of the compromised nodes by: 
\begin{equation}
\mathrm{ai}_k = \frac{\sum_{j \in C_k} \mathrm{c}_j}{N},    
\label{eq:ai}
\end{equation}
where $C_k$ is a set of compromised nodes by given attack $k$ and $N$ is the the total number of nodes. If node $j$ is already compromised, then there is no additional attack impact, $\mathrm{ai}_k=0$, introduced by attack strategy $k$. Compromising more important nodes with highly confidential information leads to early system failure (see Eq.~\eqref{eq:sf}).

\begin{table}[t]
\caption{\sc Characteristics of APT Attack Strategies}
\label{tab:AS-characteristics}
\vspace{-2mm}
\centering
\begin{tabular}{ P{0.5cm} P{1.3cm} P{1.3cm} P{1.5cm} P{2cm} }
\toprule
 $AS$ & CKC stage & Attack cost ($\mathrm{ac}$) & Node compromise & Exploited vulnerability\\
\midrule
$AS_1$ & R -- DE & 1 & No & UV \\
$AS_2$ & D -- DE & 3 & Yes (SN) & SV + EV\\
$AS_3$ & E -- DE & 3 & Yes (MN) & SV \\
$AS_4$ & E -- DE & 3 & Yes (SN) & SV + UV \\
$AS_5$ & E -- DE & 1 & Yes (SN) & UV \\
$AS_6$ & C2 -- DE & 3 & Yes (SN) & EV \\
$AS_7$ & E -- DE & 2 & Yes (SN) & EV \\
$AS_8$ & DE & 3 & Yes (SN) & S + EV \\
\bottomrule
\end{tabular}
\begin{flushleft}
Note: Each CKC stage is indicated by Reconnaissance (R), Delivery (D), Exploitation (E), Command and Control (C2), Lateral Movement (M), and Data Exfiltration (DE). Attack cost is ranged in $[1, 3]$ as an integer, representing low, medium, and high, respectively. Node compromise may involve a single node compromise (SN) or multiple nodes compromise (MN). Exploited vulnerability is indicted by Overall (O: Average vulnerability across all three types of vulnerabilities), Software (SV: software vulnerability), Encryption (EV: vulnerability by compromising encryption key(s)); and Unknown (UV: unknown vulnerability).
\end{flushleft}
\vspace{-5mm}
\end{table}

{\bf Attack Strategies}: Attackers in IoT environments have their own characteristics.  We consider several types of attacks at the different stages of the CKC by an APT attacker. The CKC consists of six stages denoted by (R, D, E, C2, M, and DE) (see Table~\ref{tab:AS-characteristics}). Each attack strategy is characterized by (1) in which CKC stage the attacker is in; (2) whether the attacker will compromise other nodes in an attack path to reach a target; (3) what attack cost $\mathrm{ac}_k$ and attack impact $\mathrm{ai}_k$) are associated with each attack strategy $k$; and (4) what vulnerability an attacker can exploit to perform a given attack strategy ($AS_k$). For simplicity, when an attacker exploits more than one vulnerability, the average security vulnerability is used to compute the normalized vulnerability, $P_i^v$. In addition, each attack strategy $k$'s attack impact, $\mathrm{ai}_k$, is obtained based on Eq.~\eqref{eq:ai}. Note that an attacker can select a non-compromised adjacent victim with the highest $APV$ value (see Eq.~\eqref{eq:attack-INT-select}) to maximize the attack success probability while minimizing the attack cost. We describe each attack strategy as follows:
\begin{itemize}
\item {\em $AS_1$ -- Monitoring attack}: This attack is to collect useful system information and identify a vulnerable node to compromise as a target. It can be performed inside or outside the network from R to DE stages. In this attack, no node compromise process is involved and accordingly its attack cost is low, $\mathrm{ac}_1 = 1$. 

\item {\em $AS_2$ -- Social engineering:}  The typical examples of this attack include email phishing, pretexting, baiting, or tailgating~\cite{Krombholz15}. We assume that an inside attacker can successfully compromise an adjacent node if the attack is successful. If the attacker is an outside attacker, it can identify a node as vulnerable during its reconnaissance stage. This attack can be performed from D to DE stages as an outside or inside attacker. Since it is highly challenging to deceive a human user who can easily detect a social engineering attack, the associated attack cost for $AS_2$ is high, $\mathrm{ac}_2 = 3$.

\item {\em $AS_3$ -- Botnet-based attack:} A botnet consists of compromised machines (or bots) running malware using C2 of a botmaster. When this attack is chosen, all compromised nodes (including original attackers) will launch epidemic attacks (e.g., spreading malware to compromise) to their adjacent, legitimate nodes~\cite{Bertino17}. This attack can be used from E to DE stages. This attack incurs high attack cost, $\mathrm{ac}_3 = 3$. 

\item {\em $AS_4$ -- Distributed Denial-of-Service (DDoS)}: A set of compromised nodes can form a botnet and perform DDoS by sending multiple requests ~\cite{Bertino17}.  When an attacker tries to compromise one of its adjacent nodes as a potential victim node, if all compromised nodes send service requests to the potential victim node, the potential victim node's vulnerability may increase because it could not properly handle all operations due to the large volume of requests received (e.g., not properly executing underlying security operations). This will allow the attacker to easily compromise the potential victim node or exfiltrate confidential data from it. To model this, unknown vulnerability, $\mathrm{uv}_i$, for a given victim node $i$ will increase for the attacker to more easily compromise a node with unknown vulnerability (e.g., increasing $\epsilon_1 \%$ for UV).  This attack can be performed from E to DE stages, with high attack cost, $\mathrm{ac}_4 = 3$. 
\item {\em $AS_5$ -- Zero-day attacks:} This attack can be performed to exploit unknown vulnerabilities of software, which are not patched yet.  The attacker can compromise chosen adjacent node $i$ based on normalized $\mathrm{uv}_j$. This attack can be performed from E to DE stages at low cost, $\mathrm{ac}_5 = 1$. 
\item {\em $AS_6$ -- Breaking encryption:} Examples include a legitimate node's private or secret key compromise. The attacker with the encryption key is considered an inside attacker with a privilege to exploit system resources. This attack can be launched from C2 to DE stages to collect system configurations or confidential information. Upon the attack success, the attacker can intercept all the information to be sent to a victim node whose private key is compromised. This attack may exploit vulnerabilities $\hat{\mathrm{ev}}_i$ associated with encryption keys and involve high attack cost, $\mathrm{ac}_6 = 3$. We assume that if a legitimate node's private key is compromised, the node is compromised. Hence, the attacker can escalate its attack by reauthenticating itself with a new password and steal confidential information or implant malware into file downloads.

\item {\em $AS_7$ -- Fake identity:} This attack can be performed when packets are transmitted without authentication or internal nodes spoofing the ID of a source node, such as MAC/IP/Virtual LAN tag spoofing in an SDN-based IoT by an SDN switch~\cite{Liu18}. This attack involves compromising a node with a fake ID. This attack can be performed from E to DE stages with cost, $\mathrm{ac}_7 = 2$. This attack increases the encryption vulnerabilities of its adjacent nodes (e.g., increasing $\epsilon_1 \%$ for EV).

\item {\em $AS_8$ -- Data exfiltration}: This attack will also allow the attacker to compromise one of the adjacent nodes. The attacker will check all data compromised by itself until DE stage. Then, if the accumulated importance of compromised data exceeds a certain threshold (i.e., $\sum_{j \in C_A} \mathrm{c}_j > \mathrm{Th}_c$), the attacker can decide whether to exfiltrate the collected intelligence to the outside.  This attack costs high with $\mathrm{ac}_8 = 3$.
\end{itemize}

We summarize the characteristics of all attack strategies considered in terms of the CKC stages involved, attack cost, node compromise, and exploited vulnerability in Table~\ref{tab:AS-characteristics}. Except $AS_1$, the attack success from $AS_2$ to $AS_8$ is determined based on whether all nodes on the attack path to reach a target node have been successfully compromised. For $AS_1$, the attack success is determined based on how long the attacker has monitored a target system. This is computed by the probability $\mathrm{vulnerability}_i \cdot e^{-1/T_A}$ where $T_A$ is the time elapsed the attacker has monitored a given target system.  This implies that the attack is likely successful when the attacker has more scanned the targeted system longer and find more vulnerabilities. After the attacker exfiltrates data successfully and leaves the system, a new attacker will arrive.  Otherwise, the attacker may be evicted by the NIDS or need to try other attack strategies to escalate its attack to a next level. 

{\bf An Attacker's Deception Detectability}: Depending on an attacker's capability, the attacker may have a different level of intelligence to detect defensive deception techniques. We denote it by $\mathrm{ad}$ to represent an attacker's probability (omitted an attacker's ID for simplicity) to detect deception used by the defender. An attacker can use this probability, $\mathrm{ad}$, to detect honeypots or honey information, as described in $DS_5$ in the next section below. 

\vspace{-3mm}
\subsection{Defense Model} \label{subsec:defense-model}

{\bf Attack Intelligence Collection}:
Different types of defense strategies can be deployed by the defender to counter APT attackers. At the same time, the NIDS will be run periodically (see Section~\ref{subsec:network-model}). Note that we don't count triggering an NIDS as one of defense strategies in order to meet a high standard of the system integrity.  When an attacker arrives at the system as an inside attacker (i.e., after the E stage), it can be detected by the NIDS. However, the system aims to collect more attack intelligence (e.g., attack signatures), which can improve the NIDS as a long-term goal. Thus, depending on the perceived risk level from the attacker, the system will determine whether to keep the detected attacker in the system or evict it.  We estimate the perceived system risk level based on the criticality level of the compromised node, $\mathrm{c}_i$, and determine if the system will allow the attacker to reside in the system or be evicted according to predefined risk threshold, $\mathrm{Th_{risk}} \in [0, 1]$. The decision to evict node $i$, which is detected as compromised, can be given by:
\begin{eqnarray}
\mathrm{Evict}_i =
\begin{cases}
1\; \; \text{if} \; \; \mathrm{c}_i > \mathrm{Th_{risk}} \\
0 \; \; \text{otherwise.}
\end{cases}
\label{eq:detect-risk}
\end{eqnarray}
Here $\mathrm{Evict}_i =1$ means evicting node $i$ while $\mathrm{Evict}_i =0$ means allowing node $i$ to reside in the system. Note that this rule is applied when node $i$ is detected as compromised by the NIDS regardless of its correctness. Hence, false positive nodes can be also assessed by this rule while false negative nodes can safely reside in the system without being assessed by $\mathrm{Th_{risk}}$. 

When nodes detected as compromised (i.e., true and false positives) are evicted, all associated edges will be disconnected, which may generate some non-compromised nodes being isolated from the network. To maintain connectivity of non-compromised but isolated nodes, we connect them to the network based on $P_i^r$ to maintain node $i$'s mean degree based on the ER network model~\cite{Newman10}.  To deal with the attackers (or compromised nodes) residing in the system, which are either false negatives or attackers kept to collect further attack intelligence, the defender system can take the several defense strategies. Each strategy $k$ will be represented by: (i) defense cost ($\mathrm{dc}_k$) in time/complexity and expense, where $\mathrm{dc}_k \in [0, 3]$ as an integer for no, low, medium, and high cost, respectively; (ii) defense impact ($\mathrm{di}_k$) for its defense effectiveness; (iii) the stage of the CKC (i.e., R, D, E, C2, LM, or DE) for strategy $k$ being used; and (iv) system change on what actual changes are made in the system (e.g., what vulnerabilities are reduced or network topology or cryptographic keys being changed). The defense impact, $\mathrm{di}_k$, is computed by:
\begin{eqnarray}
\mathrm{di}_k = 1-\mathrm{ai}_k,
\label{eq:di}
\end{eqnarray}
where $\mathrm{ai}_k$ is the attack impact introduced by strategy $k$ in Eq.~\eqref{eq:ai}. We measure the effectiveness of a defense strategy as the opposite impact of attack success (i.e., successfully compromising a node). That is, attack failure will increase the impact of the defense strategy. 

{\bf Defense Strategies}: This work considers the following defense strategies:
\begin{itemize}
\item {\em ${DS}_1$ -- Firewalls}: We assume that firewalls are implemented in the SDN controller to monitor and control the incoming and outgoing packet flows according to predefined rules. We model the effectiveness of firewalls by lowering down unknown vulnerabilities ($\mathrm{uv}_i$) all over the network. Specifically, firewall is assumed to reduce vulnerabilities to outside attackers by a certain percent (i.e., $\epsilon_2 \%$). 

\item {\em $DS_2$ -- Patch Management}: Known vulnerabilities can be patched by a given defense system~\cite{Leiba18-patches}. A patch is used to temporarily fix software vulnerabilities or provide updates in a full software package. A patch refers to a software update such as code to be installed in a software program. This will decrease software vulnerabilities ($\mathrm{sv}_i$) of all nodes, such as decreasing a certain percent of the vulnerability (i.e., $\epsilon_2 \%$).
\item {\em $DS_3$ -- Rekeying Cryptographic Keys}: Cryptographic keys used for all nodes in the network are rekeyed, which lowers the encryption vulnerability by setting $\mathrm{T_{rekey}} = 1$ which reduces $\hat{\mathrm{ev}}_i = \mathrm{ev}_i \cdot e^{-1/\mathrm{T_{rekey}}}$.
\item {\em $DS_4$ -- Eviction}: Recall that an attacker with low risk (see Eq.~\eqref{eq:detect-risk}) is allowed to stay in the system for collecting attack intelligence. However, as the system is at risk due to high security vulnerability in terms of the amount of compromised confidential information (i.e., importance; see Eq.~\eqref{eq:sf}), all inside attackers (or compromised nodes) will be evicted from the system.  However, the false negatives will remain in the system while a substantial number of compromised nodes is evicted using $DS_4$. 

\item {\em $DS_5$ -- Low/high-interaction honeypots (LHs/HHs)}~\cite{Kyung17-honeypots}: LHs and HHs can be activated as a defense strategy. LHs and HHs differ in their deception detectability and cost. In a given network, we deploy a set of LHs and HHs which are deactivated in the deployment phase. When this strategy is selected, they will be activated, which will change the network topology as LHs and HHs are to be connected with a number of nodes in the network. Hence, $DS_5$ will change attack paths and lure attackers to the honeypots. To be specific, when $DS_5$ is selected, LHs and HHs will be activated. This will enable them to be connected to highly vulnerable nodes based on $\mathrm{vulnerability}_i$ where HHs will be connected to nodes of higher vulnerability than nodes connected to LHs. In order for the attacker not to reach legitimate nodes, we will only allow incoming connections (i.e., in-degree) from legitimate nodes to the honeypots. Once the attacker is caught by one of the implemented honeypots, it will be diverted to a fake network for monitoring purposes. Recall that an attacker can detect the deception with $\mathrm{ad}$ for a LH and $\mathrm{ad}/2$ for a HH.

\item {\em $DS_6$ -- Honey information}: This defense strategy can lure attackers by disseminating false information, such as honey token, fake patch, honey files, or bait files.  This strategy will involve the dissemination of false system vulnerability information, such as providing high (low) vulnerabilities for less (more) vulnerable nodes. The attacker will need to detect whether a known vulnerability of a potential victim node is true or fake according to its deception detectability, $\mathrm{ad}$. If the attacker is successfully deceived, it will make an attack strategy decision based on incorrect vulnerability information.

\item {\em $DS_7$ -- Fake keys}~\cite{almeshekah2016cyber}: Fake keys can be planted for potential, inside attackers which may use a fake key obtained by compromising another legitimate, inside node to communicate with other nodes to obtain more confidential information. This will be realized that even if the attacker compromises a cryptographic key (e.g., $AS_2, AS_6, AS_7, AS_8$), a potential victim targeted by the attacker may not be compromised. We model this using the probability the attacker obtains a fake key implanted in nodes, $P_{fake}$. When the attacker obtains the fake key of a node, the node will not be compromised. 

\item {\em $DS_8$ -- Hiding network topology edges}: This strategy hides $c_{NT}$\% of network edges in order to hide an actual network topology to an attacker. We use a simple rule for each node to hide the edge with the most critical adjacent node based on its criticality value, $\mathrm{c}_i$.
\end{itemize}

\begin{table}[t]
\footnotesize
\vspace{-3mm}
\caption{\sc Characteristics of Defense Strategies}
\label{tab:DS-characteristics}
\vspace{-3mm}
\centering
\begin{tabular}{ P{0.5cm} P{1.3cm} P{1.3cm} P{3.5cm} }
\toprule
 $DS$ & CKC stage & Defense cost ($\mathrm{dc}$) & System change ($\mathrm{dsc}$)\\
\midrule
$DS_1$ & R -- D & 1 & Lowering UV \\
$DS_2$ & D -- DE & 2 & Lowering SV \\
$DS_3$ & E -- DE & 3 & Lowering EV \\
$DS_4$ & E -- DE & 3 & Evict all compromised nodes \\
$DS_5$ & E -- DE & 3 & Lure attackers to with LHs and HHs \\
$DS_6$ & C2 -- DE & 1 & Disseminate fake system vulnerability information \\
$DS_7$ & E -- DE & 2 & Plant a fake key \\
$DS_8$ & R -- DE & 2 & Hide critical network edges \\
\bottomrule
\end{tabular}
\begin{flushleft}
Note: Each CKC stage is indicated by Reconnaissance (R), Delivery (D), Exploitation (E), Command and Control (C2), Lateral Movement (M), and Data Exfiltration (DE). Defense cost is ranged in $[1, 3]$ as an integer, representing low, medium, and high, respectively. System change may involve lowering unknown vulnerabilities (UV), software vulnerabilities (SV), or encryption vulnerabilities (EV).
\end{flushleft}
\vspace{-10mm}
\end{table}

All defense strategies will have corresponding defense costs ($\mathrm{dc}_k$'s) and are believed useful when the attacker are in certain CKC stages based on the defender's belief. This is used for the defender to choose each subgame based on hypergame theory.  We summarized the characteristics of each defense strategy considered in Table~\ref{tab:DS-characteristics}.

{\bf System Failure Conditions}: We define that a system failure (SF) occurs when the following condition is met:
\begin{equation}
SF = 
\begin{cases}
1 \; \; \text{if}\; \; \rho_1 \leq \frac{\sum_{i \in G} \mathrm{cp}_i \cdot \mathrm{Importance}_i}{\sum_{i \in G} \mathrm{Importance}_i} \; || \; \rho_2 \geq \frac{|G_t|}{|G|} \\
0 \; \; \text{otherwise.}
\end{cases}
\label{eq:sf}    
\end{equation}
Here $G_t$ refers to a network at time $t$ which does not include nodes being evicted while $G$ is an original network. Hence $|G|$ and $|G_t|$ are the number of the original nodes and the number of the current nodes in the system at time $t$, respectively.  $\rho_1$ is a threshold as a fraction to determine whether a system fails or not based on the sum of compromised nodes' importance values over the sum of all nodes' importance values. SF mainly captures the system failure caused by the loss of three security goals, such as confidentiality, integrity, and availability.  $\rho_2$ is a threshold to determine whether a system can functionally operate based on a sufficient number of active nodes at time $t$.

\section{Attack-Defense Hypergame} \label{sec:case_study}

First, the attacker will select strategy $AS_1$ to monitor a target system in the reconnaissance (R) stage, aiming to penetrate into it as a legitimate user. If the attack is successful based on the success probability $\mathrm{vulnerability}_i \cdot e^{-1/T_A}$, the attacker can proceed to the delivery (D) stage of the CKC. In the D stage, the attacker can choose one of the two strategies $AS_1$ and $AS_2$. If the attacker can successfully compromise a targeted victim node, which is one of its adjacent nodes, it can successfully penetrate the system and become an inside attacker with legitimate credentials. Now the attacker is in the Exploitation (E) stage. From E to data exfiltration (DE) stages, any inside attacker detected can be assessed by the defender on whether it can stay in the system based on the risk assessment in Eq.~\eqref{eq:detect-risk}. Hence, depending on the criticality of the attacked node, the attacker can be detected by the NIDS or be kept in the system if the defense system intends to collect attack intelligence from it.  To assess such risk, the attacker should be detected as an attacker (i.e., true and false positives) by the NIDS. If not (i.e., false negatives), the attacker can safely stay even without being detected. From E to DE, the attack is determined as successful if $\mathrm{ai}_i > 0$ (see Eq.~\eqref{eq:ai}).  If the original attacker (i.e., a node the attacker is on) is evicted, then a new attacker will arrive. If an attacker is successful by taking $AS_8$ (data exfiltrated), it will leave the system and a new attacker will arrive. This process will continue until the system fails based on Eq.~\eqref{eq:sf}.

Next we formulate the hypergame between the attacker and defender, and define the game components.  We provided the detailed explanation of hypergame theory formulations and its related equations in Appendix A, which are used in the sections below.
\vspace{-2mm}
\subsection{Utilities} \label{subsec:utilities}


{\bf An attacker's utility} ($u_{pq}^A$) corresponding to attack strategy $p$ ($AS_p$) can be expressed as the difference between attack gain and attack loss. The attacker's utility ($u_{pq}^A$) when the attacker takes $AS_p$ and the defender takes $DS_q$ is calculated by:
\begin{gather}
u_{pq}^A =  G_{pq}^A - L_{pq}^A, \; \;
G_{pq}^A = \mathrm{ai}_p + \mathrm{dc}_q, \; \; L_{pq}^A = \mathrm{ac}_p + \mathrm{di}_q,
\label{eq:attack-utility}    
\end{gather}
where the attack and defense cost (i.e., $\mathrm{ac}_p$ and $\mathrm{dc}_q$) and the attack and defense impact (i.e., $\mathrm{ai}_p$ and $\mathrm{di}_q$) are discussed in Sections~\ref{subsec:attack-model} and \ref{subsec:defense-model}, respectively.

{\bf A defender's utility} ($u_{qp}^D$) by selecting $DS_q$ when the attacker takes $AS_p$ can be computed based on the difference between the gain and loss by:   
\begin{gather}
u_{qp}^D =  G_{qp}^D - L_{qp}^D,\; \;
G_{qp}^D = \mathrm{di}_q + \mathrm{ac}_p, \; \; L_{qp}^D = \mathrm{dc}_q + \mathrm{ai}_p.
\label{eq:defense-utility}    
\end{gather}
Similar to $u_{pq}^A$, the attack and defense cost (i.e., $\mathrm{dc}_q$ and $\mathrm{ac}_p$) and the attack and defense impact (i.e., $\mathrm{di}_q$ and $\mathrm{ai}_p$) are computed. We consider a zero-sum game between the attacker and defender (i.e., $u_{pq}^A + u_{qp}^D=0$).

\begin{table}[t]
\centering
\vspace{-2mm}
\caption{\sc Possible Strategies Under Each Stage of the CKC}
\label{tab:subgame}
\vspace{-3mm}
\begin{tabular}{ P{1cm} P{1.5cm} P{2.2cm} P{2.4cm} }
\toprule
Subgame & CKC stage & Attack strategies & Defense strategies \\
\midrule
0 & Full game & $AS_1-AS_8$ & $DS_1-DS_8$ \\
1 & R & $AS_1$ & $DS_1, DS_8$  \\
2 & D & $AS_1, AS_2$ & $DS_1, DS_2$  \\
3 & E & $AS_1-AS_5$, $AS_7$ & $DS_3-DS_5, DS_7$  \\
4 & C2 & $AS_1$--$AS_7$ & $DS_3$--$DS_8$  \\
5 & M & $AS_1$--$AS_7$ & $DS_3$--$DS_8$  \\
6 & DE & $AS_1$--$AS_8$ & $DS_3$--$DS_8$  \\
\bottomrule
\end{tabular}
\vspace{-5mm}
\end{table}

\vspace{-2mm}
\subsection{Estimation of Uncertainty} \label{subsec:uncertainty-estimation}

As in Eq.~\eqref{eq:heu} in Appendix A, an attacker's and defender's hypergame expected utilities (HEUs) are estimated based on the level of uncertainty, $g$, perceived by each player. In this section, we show how the level of $g$ is estimated by the attacker (i.e., $g^A$) and the defender  (i.e., $g^D$). Note that we omit an ID of the attacker and defender for simplicity. 

We model {\bf an attacker's perceived uncertainty} based whether a defensive deception is used and how long the attacker has monitored a target system.  That is, given the time period the attacker has monitored in a target system ($T_A$) and a defense strategy taken ($\mathrm{df}$), the attacker's uncertainty ($g^A$) is estimated by: 
\begin{eqnarray}
g^A  =  1- \exp(-\lambda \cdot \mathrm{df}/T_A). \label{eq:attack_uncertainty}
\end{eqnarray}
Here $\lambda$ is a parameter of representing an amount of initial knowledge towards a given system configuration (higher $\lambda$ increases uncertainty, and vice versa) and $\mathrm{df} = 1 + (1-\mathrm{ad}) \cdot \mathrm{dec}$.  $\mathrm{df}$ returns 1 when no defensive deception is used (i.e., $\mathrm{dec} =0$); it returns $1 + (1 - \mathrm{ad})$ where $\mathrm{ad}$ refers to an attacker's deception detectability in $[0, 1]$ when defensive deception is used (i.e., $\mathrm{dec} =\mathrm{dc}$ where $\mathrm{dc}$ is defense cost implying that higher defense cost allows higher quality of deception). The formulation of $g^A$ above implies that the attacker has lower uncertainty as it has monitored the target system longer. On the other hand, the attacker has higher uncertainty when it has lower deception detectability and the defender uses a defensive deception strategy. Hence, we set $\mathrm{dec} = \mathrm{dc}$ (defense cost) when defensive deception strategies, $DS_5$--$DS_8$, are taken while setting $\mathrm{dec} =0$ when non-deception defense strategies, $DS_1 - DS_4$, are taken.

{\bf A defender's uncertainty} towards an attacker increases as it has monitored the attacker for a longer period where the defender's monitoring time towards the attacker is denoted by $T_D$. In addition, if the attacker has not been deceived by defense strategies, it is assumed to be intelligent not to expose its information to the defender. Considering these two, we model $g^D$ by:
\begin{eqnarray}
g^D  =  1-\exp(-\mu \cdot \mathrm{ad}/T_D),
\label{eq:defense_uncertainty}
\end{eqnarray}
where $\mu$ is a parameter of representing an amount of initial knowledge towards an attacker (higher $\lambda$ increases uncertainty, and vice versa), $\mathrm{ad}$ is an attacker's deception detectability, and $T_D$ is a defender's accumulated monitoring time towards the attacker.  In $g^D$, the defender perceives lower uncertainty at longer $T_D$ while perceiving higher uncertainty at higher $\mathrm{ad}$.


\vspace{-2mm}
\subsection{Estimation of HEUs} \label{subsec:heu_estimation_example}

In order to calculate the HEU for each player (see Eq.~\eqref{eq:heu} in Appendix A), we need to obtain $P_\kappa$ (i.e., the probability a row player chooses subgame $\kappa$), $r_{\kappa p}$ (i.e., the probability that a row player takes strategy $k$ in subgame $\kappa$), and $c_{\kappa j}$ (i.e., the probability that a column player takes strategy $h$ in subgame $\kappa$ based on a row player's belief) because $S_q$ is estimated based on $P_\kappa$ and $c_{\kappa h}$ while $r_{\kappa p}$ is needed when a row player considers strategy $k$. 

\textbf{Computation of $P_\kappa$}: Recall that $P_\kappa$ refers to the probability that subgame $\kappa$ is played by a row player. We notate this for an attacker and a defender by $P_\kappa^A$ and $P_\kappa^D$, respectively. We define a subgame based on where an attacker is located in the stages of the CKC which will determine a set of available strategies for both parties. We assume that the attacker clearly knows where it is located in the CKC while the defender is not certain about the stage of the attacker in the CKC. We model the defender's $P_\kappa^D$ based on its uncertainty $g^D$. Thus, the defender can know the CKC stage of the attacker with $1-g^D$ (certainty) and correctly choose a subgame based on the attacker's actual stage in the CKC. With $g^D$, the defender will choose subgame $0$ (i.e., a full game with all available strategies). The set of available strategies may be different depending on what subgame to play, as shown in Table~\ref{tab:subgame}.

\textbf{Computation of $r_{\kappa h}$ and $c_{\kappa h}$}: $r_{\kappa h}$ is the probability that a row player will play strategy $h$. We denote this for the attacker and defender by $r_{\kappa p}^A$ and $r_{\kappa q}^D$ for attack strategy $p$ and defense strategy $q$, respectively. $c_{\kappa h}$ is the probability that a column player will take strategy $h$ based on a row player's belief.  We also denote this for the attacker and defender by $c_{\kappa p}^A$ and $c_{\kappa q}^D$ attack strategy $p$ and defense strategy $q$, respectively. In the very beginning, since no historical information is available, each player will use a uniform probability by choosing one of available strategies in a chosen subgame with an equal probability, meaning choosing a strategy at random. As players participate in repeated games, their recorded history regarding what strategies have been taken is available. Then, we will use Dirichlet distribution~\cite{Teh10-dirichlet} to model multinomial probabilities based on the strategies taken for past repeated games. If either an attacker or defender is certain about the opponent's strategy, it will estimate its corresponding $r_{\kappa p}^A$, $r_{\kappa q}^D$, $c_{\kappa p}^A$, and $c_{\kappa q}^D$ as:  
\vspace{-2mm}
\begin{gather}
\footnotesize
r_{\kappa p}^A = \frac{\gamma_{\kappa p}^A}{\sum_{p \in \mathbf{AS}_{\kappa}} \gamma_p^A}, \; \; 
c_{\kappa q}^A = \frac{\gamma_{\kappa q}^D}{\sum_{q \in \mathbf{DS}_{\kappa}} \gamma_q^D}, \; \; \\
r_{\kappa q}^D = \frac{\gamma_{\kappa q}^D}{\sum_{q \in \mathbf{DS}_{\kappa}} \gamma_q^D}, \; \;
c_{\kappa p}^D = \frac{\gamma_{\kappa p}^A}{\sum_{p \in \mathbf{AS}_{\kappa}} \gamma_p^A}.
\label{eq:c-kappa}  
\vspace{-2mm}
\end{gather}
Note that $\mathbf{AS}_{\kappa}$ and $\mathbf{DS}_{\kappa}$ are a set of attack strategies and defense strategies, respectively. $\gamma_{q}^D$ ($\gamma_{p}^A$) is the number of times the defender (attacker) will take strategy $q$ (or $p$) based on the attacker's (defender's) belief up to time $(t-1)$ where the current state is at time $t$. Since the probability of a column player playing a particular strategy is estimated by a row player's belief, ground truth $c_{\kappa q}^A$ and $c_{\kappa p}^D$ (as shown in the equations above) will be only detected with the probability $(1-g^A)$ and $(1-g^D)$ when the row player is an attacker or a defender, respectively. Otherwise, the row player will select one among the available strategies in a given subgame $\kappa$ at random due to the uncertainty.

{\bf An attacker's HEU (AHEU)} is computed with: (1) attack utilities (i.e., $u_{kh}^A$'s in Eq.~\eqref{eq:attack-utility}); (2) the attacker's belief about defense strategy $h$ (i.e., $S_q^A$ in Eq.~\eqref{eq:S_i} in Appendix A); and (3) the attacker's perceived uncertainty (i.e., $g^A$ in Eq.~\eqref{eq:attack_uncertainty}). 

Similarly, {\bf a defender's HEU (DHEU)} is estimated using: (1) defense utilities (i.e., $u_{qp}^D$'s in Eq.~\eqref{eq:defense-utility}); (2) the defender's belief about attack strategy $p$ (i.e., $S_p^D$ in Eq.~\eqref{eq:S_i} in Appendix A); and (3) the defender's perceived uncertainty (i.e., $g^D$ in Eq.~\eqref{eq:defense_uncertainty}). Both AHEU and DHEU can be obtained based on Eq.~\eqref{eq:heu} in Appendix A.  Since the row player selects each strategy $h$ based on $r_{\kappa h}$ for given subgame $\kappa$, we calculate AHEU and DHEU as follows:  
\begin{gather}
\text{AHEU} (rs_p^A, g^A) = \text{HEU} (rs_p^A, g^A), 
\nonumber \\
\text{DHEU} (rs_q^D, g^D) = \text{HEU} (rs_q^D, g^D).
\label{eq:c-AHEU-DHEU}    
\end{gather}
A player will play a strategy according to the probability distribution of strategies available in a given subgame $\kappa$.

\section{Experimental Setting}
\label{sec:experimental-setting}

In this work, we use the following {\bf metrics}: 
\begin{itemize}
\item {\em Perceived Uncertainty Level} ($g^A$ or $g^D$): An attacker's or a defender's mean uncertainty level which is measured as shown in Eqs.~\eqref{eq:attack_uncertainty} and~\eqref{eq:defense_uncertainty}, respectively.
\item {\em Hypergame Expected Utility} (HEU):  This metric measures the HEU of played strategies profile, according to Eq.~\eqref{eq:heu} in Appendix A in the supplement document.
\item {\em Cost for Taking a Chosen Strategy} ($C_A$ or $C_D$): This metric measures the average attack (or defense) cost paid by an attacker (or a defender) to play a specific strategy. Attack cost ($C_A$) and defense cost ($C_D$) of all available strategies are summarized in Tables~\ref{tab:AS-characteristics} and \ref{tab:DS-characteristics}, respectively. For a given scenario consisting of a series of games until the system fails based on Eq.~\eqref{eq:sf}, the average attack or defense cost per game is demonstrated. 
\item {\em Mean Time to Security Failure} (MTTSF): This metric measures a system lifetime based on the system states that do not fall in the system failure states based on Eq.~\eqref{eq:sf}.
\item {\em TPR of an NIDS}: This metric measures the true positive rate of the NIDS in order to observe how much defensive deception can improve the quality of the NIDS based on the attack intelligence collected during the time of using defensive deception.
\end{itemize}

Our work compares the performance following schemes:
\begin{itemize}
\item {\em Game with defensive deception and perfect information (DD-PI)}: This scheme plays a game where each player has perfect information regarding which strategy is played by its opponent, which means there is no uncertainty, $g=0$ (i.e., $g^A=g^D=0$), when a defender uses all defensive deception (DD) strategies. 

\item {\em Game without defensive deception and perfect information (No-DD-PI)}: This scheme plays a game where each player has perfect information regarding what strategy its opponent plays (i.e., $g=0$) when the defender does not use DD strategies. 
\item {\em Hypergame with defensive deception and imperfect information (DD-IPI)}: This scheme plays a game where each player does not have perfect information regarding the strategy of its opponent (i.e., $g>0$ with $g_A > 0, g^D > 0$) when the defender uses DD strategies. This is our proposed scheme that considers uncertainty $g$ (i.e., imperfect information, IPI) and DD.

\item {\em Hypergame without defensive deception and imperfect information (No-DD-IPI)}: This scheme plays a game where each player does not have perfect information towards what strategy its opponent takes (i.e., $g>0$) when the defender does not use DD strategies. 
\end{itemize}

We consider 500 nodes in a given network where a network topology is generated by the ER random graph model with $G(N, P^r)$ where $N$ is the total number of nodes and $ P^r (= P_i^r)$ is the connection probability between any pair of nodes~\cite{Newman10}. To consider honeypots with low or high interactions, we also assign 75 nodes as honeypots with 50 LHs and 25 HHs. For honeypots, we maintain a directed network where the outgoing edges (i.e., out-degree) are from each honeypot to all other honeypots to ensure an attacker not to be connected with other legitimate nodes. When a honeypot is activated (i.e., $DS_5$), highly vulnerable nodes are connected to the honeypot as an incoming edge (i.e., in-degree). However, outgoing edges from the honeypot are always forwarded to other honeypots, not real legitimate nodes, which are protected from the attacker. In our experiment, when the honeypots are activated, the top 225 vulnerable nodes are connected to honeypots where top 75 vulnerable nodes are connected to 25 HHs and next top 150 vulnerable nodes are connected to 50 LHs.  We assume that the defender has inherently higher uncertainty regarding an attacker while the attacker has a certain level of knowledge regarding a system due to its reconnaissance effort before becoming an inside attacker.  This was reflected by setting $\lambda=0.8$ and $\mu=8$ in Eqs.~\eqref{eq:attack_uncertainty} and~\eqref{eq:defense_uncertainty}, respectively. We summarized the notations of key design parameters, their meaning, and default values used in Table \ref{tab:notations-table} in Appendix B of the submitted supplement document.

\begin{figure*}
\centering
\subfloat{\includegraphics[width=.12\textwidth]{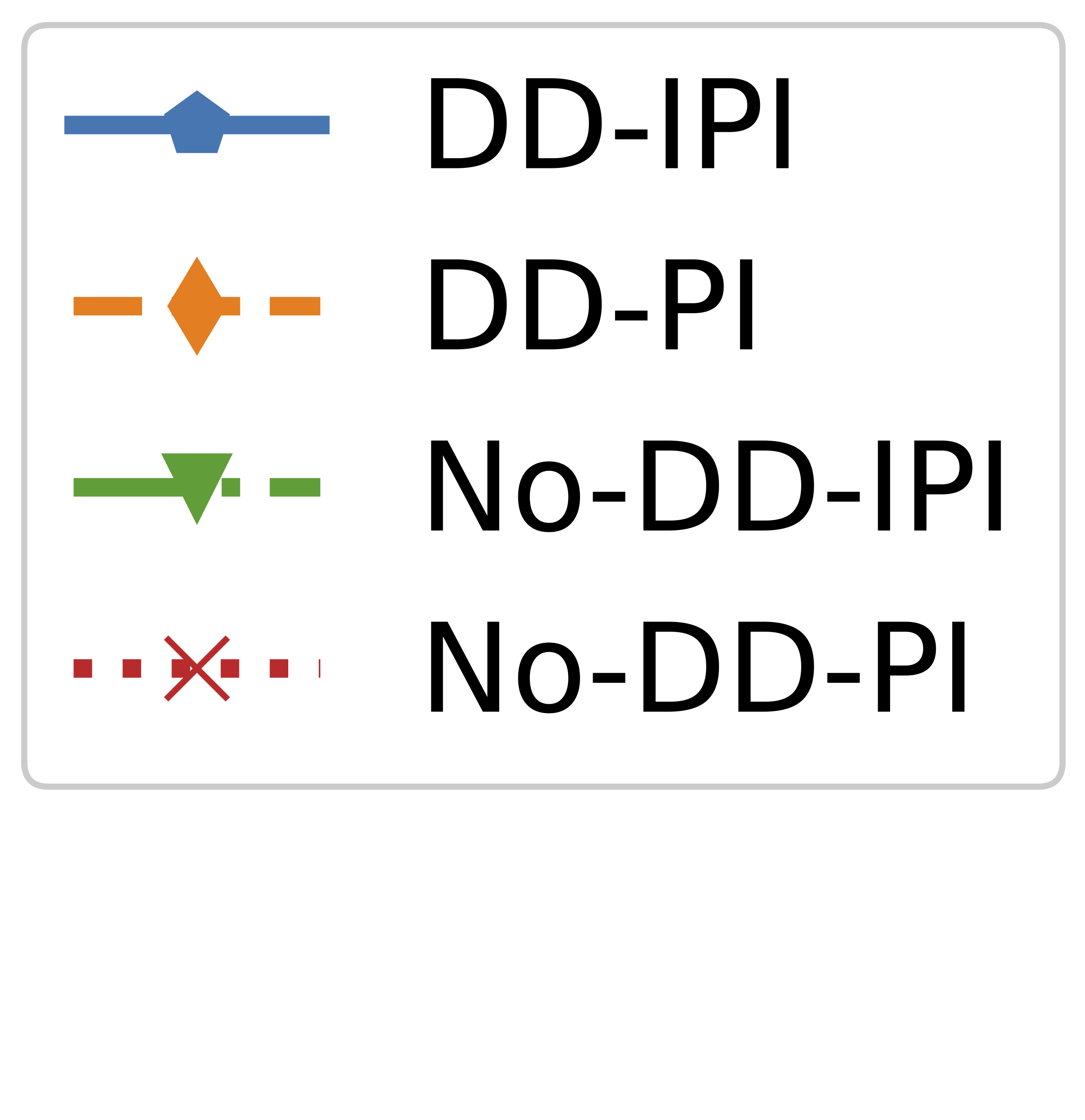}\label{fig: attack-legend}} 
 \setcounter{subfigure}{0}
\hfil
\subfloat[Attacker's uncertainty]{\includegraphics[width=.29\textwidth]{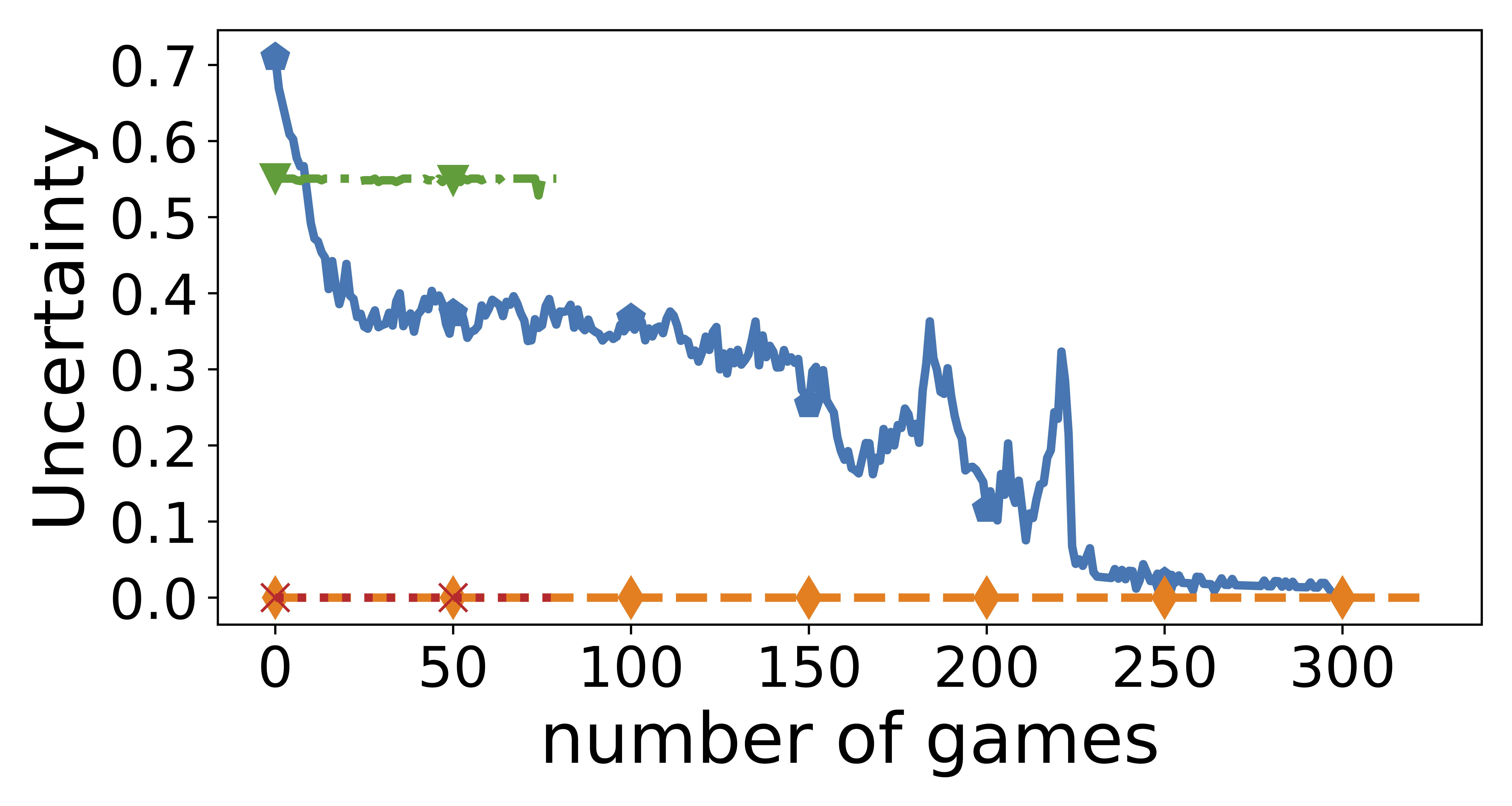}\label{fig: attack-uncertainty}} 
\hfil
\subfloat[Attacker's HEU]{\includegraphics[width=.29\textwidth]{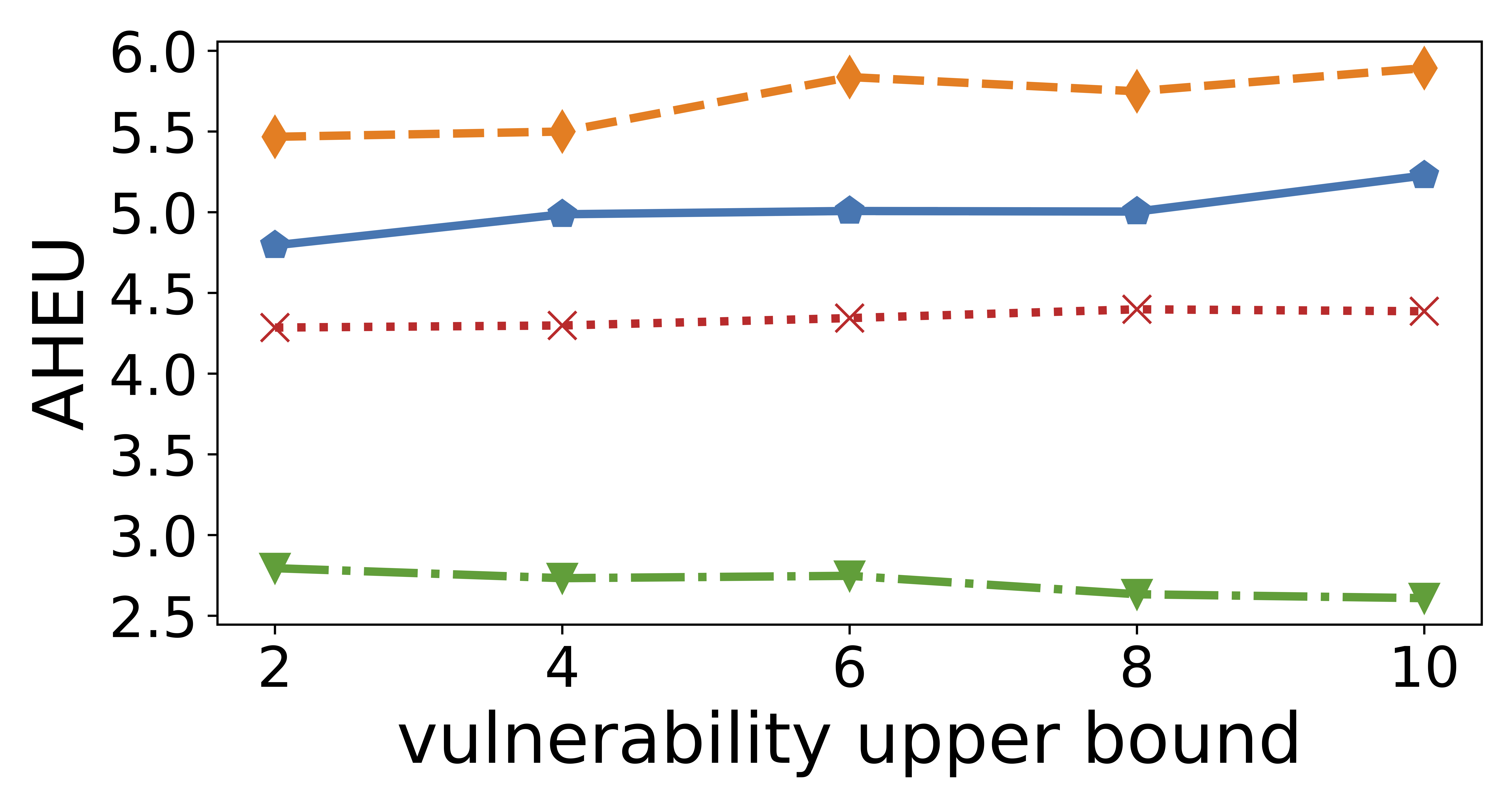}\label{fig: attack-heu}} 
\hfil
\subfloat[Attack cost]{\includegraphics[width=.29\textwidth]{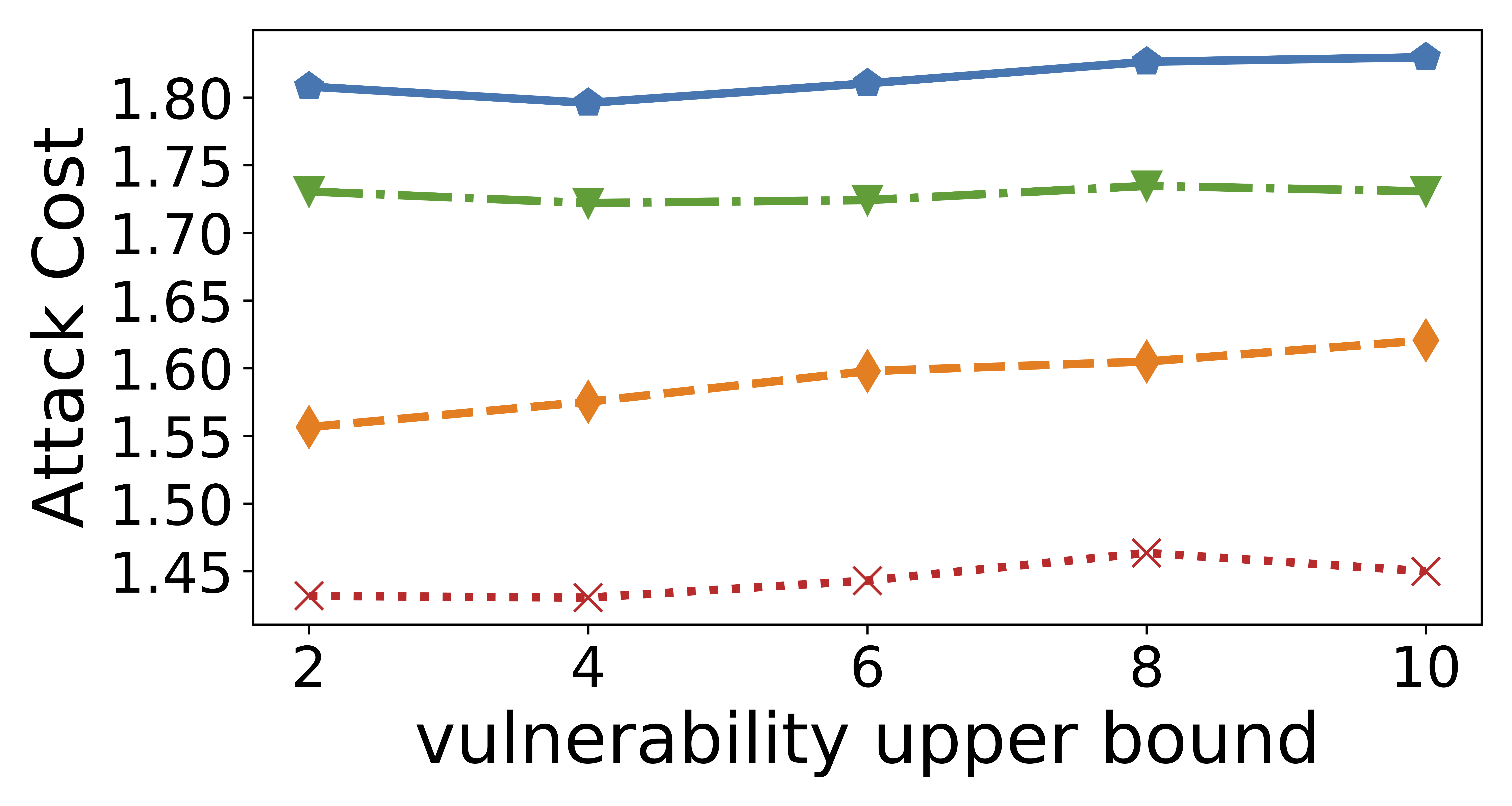}\label{fig: attack-cost}} 
\hfil    
\caption{An attacker's uncertainty, hyergame expected utility (AHEU), and attack cost. The `vulnerability upper bound' ($U_v$) refers to the CVSS-based software vulnerability score of IoT devices, Web servers and Databases, which is scaled in $[1, U_v]$.}
\label{fig:attacker}
\vspace{-3mm}
\end{figure*}

\begin{figure*}
\vspace{-3mm}
\centering
\subfloat{\includegraphics[width=.12\textwidth]{./figs/FINAL/legend_1.png}\label{fig: defender-legend}} 
 \setcounter{subfigure}{0}
\hfil    
\subfloat[Defender's uncertainty]{\includegraphics[width=.29\textwidth]{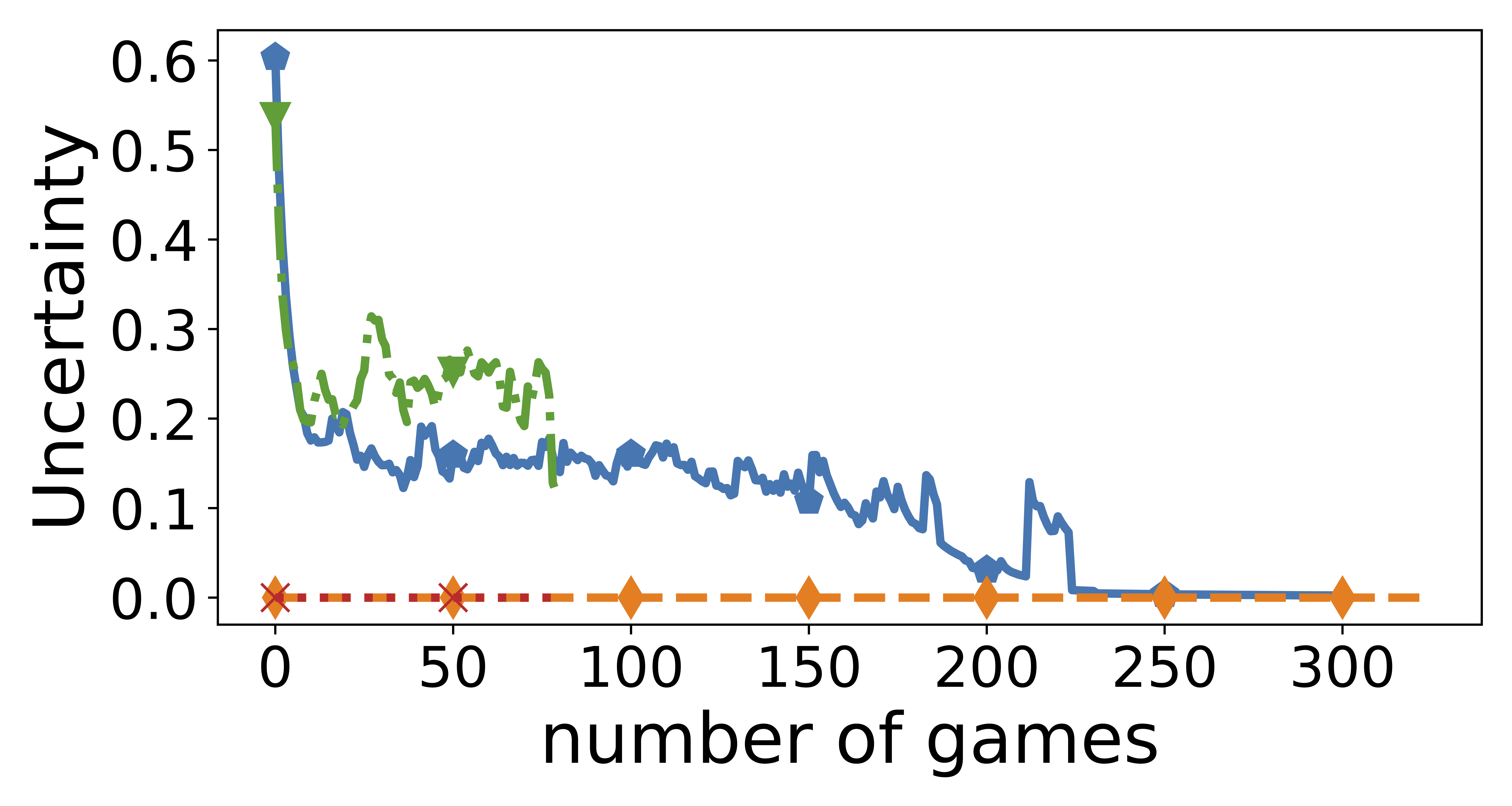}\label{fig: def-uncertainty}} 
\hfil    
\subfloat[Defender's HEU]{\includegraphics[width=.29\textwidth]{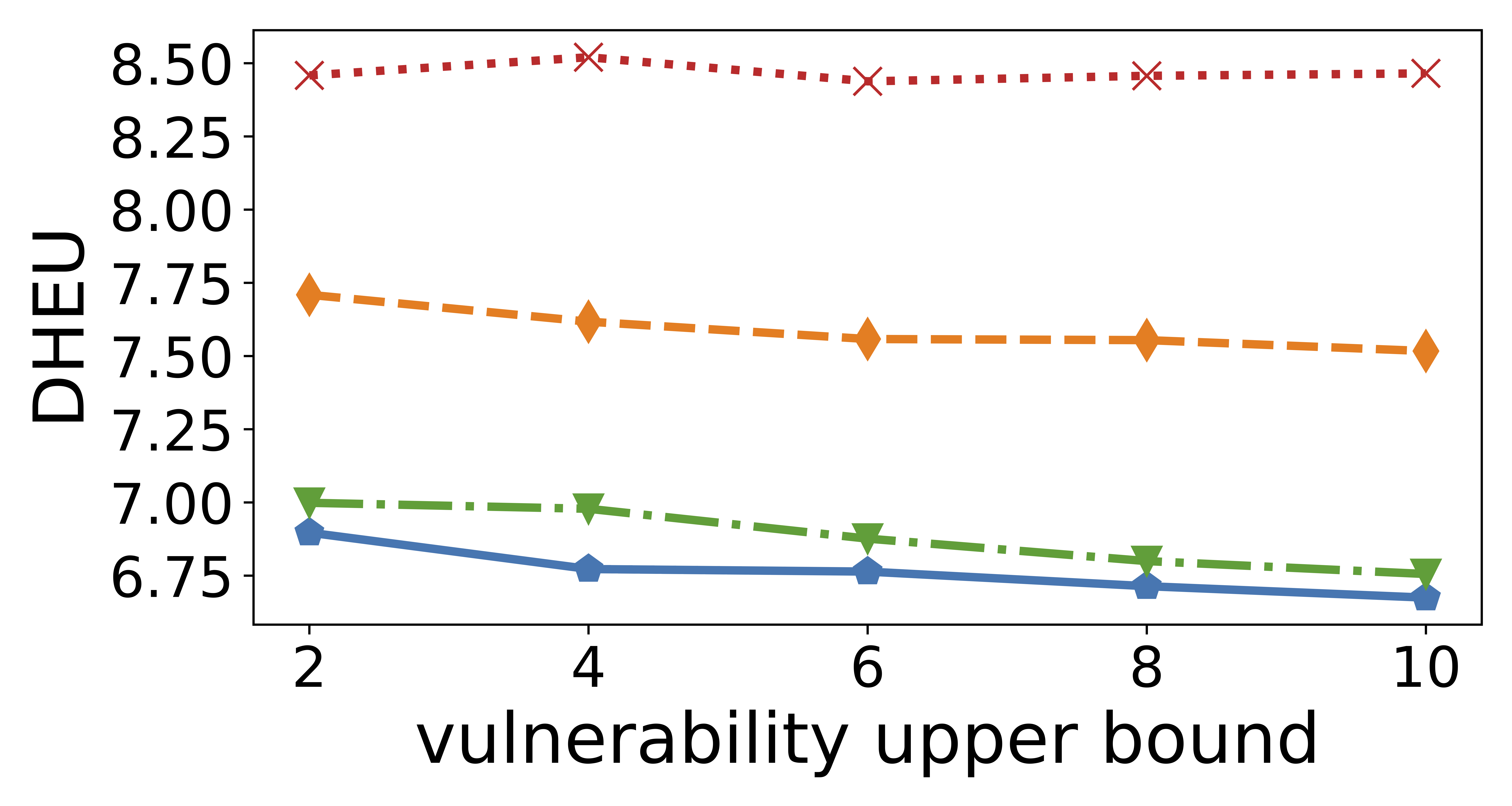}\label{fig: def-heu}} 
\hfil
\subfloat[Defense cost]{\includegraphics[width=.29\textwidth]{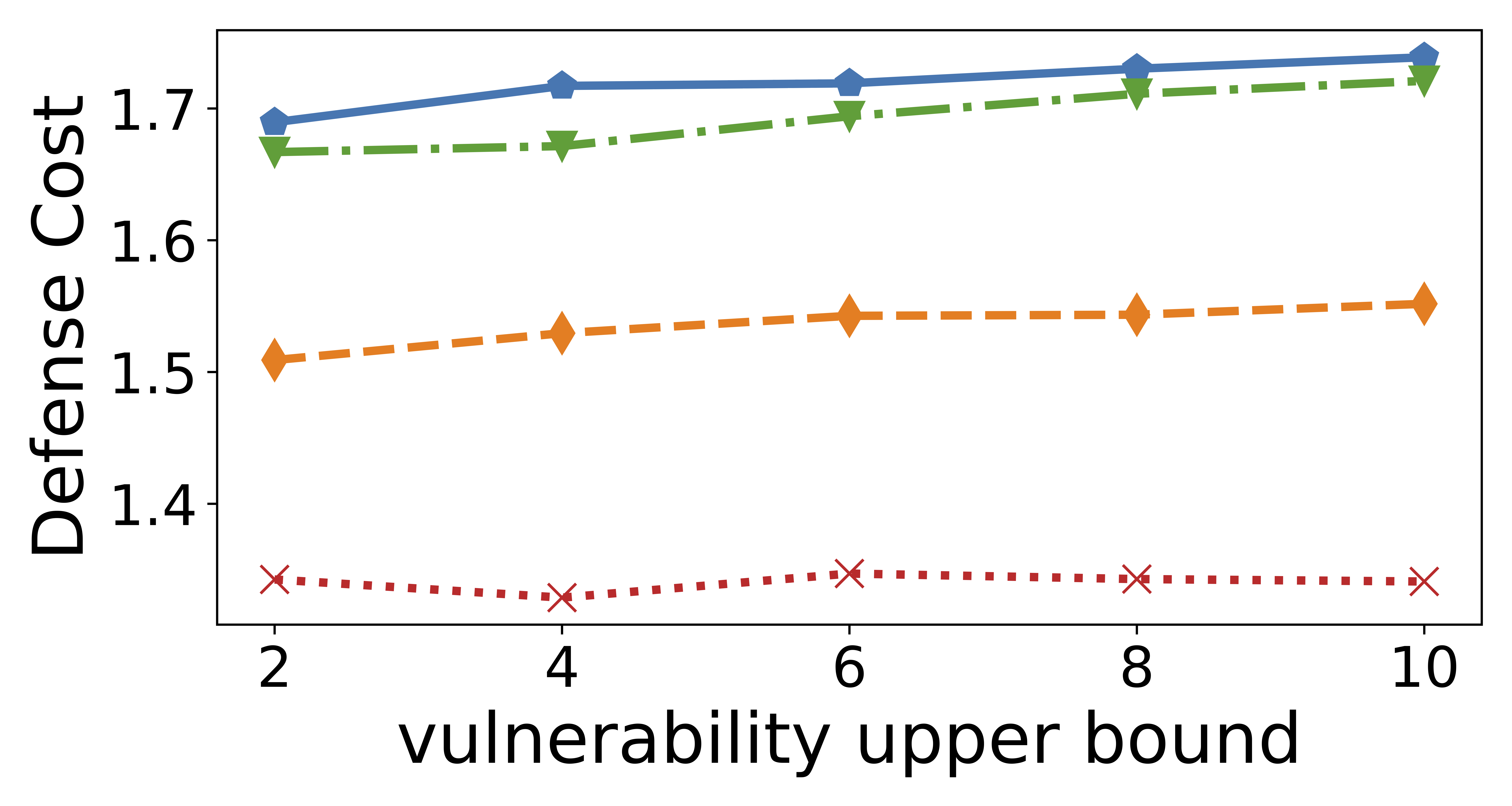}\label{fig: def-cost}} 
\caption{A defender's uncertainty, hyergame expected utility (DHEU), and defense cost. The `vulnerability upper bound' ($U_v$) refers to the CVSS-based software vulnerability score of IoT devices, Web servers and Databases, which is scaled in $[1, U_v]$.}
\label{fig:defender}
\vspace{-3mm}
\end{figure*}

\begin{figure}
\centering
\vspace{-3mm}
\subfloat{\includegraphics[width=.48\textwidth]{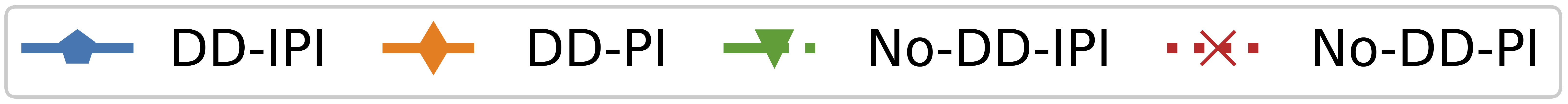}} 
\vspace{-4mm}
\hfil
\subfloat[MTTSF]{\includegraphics[width=.24\textwidth, height=0.2\textwidth]{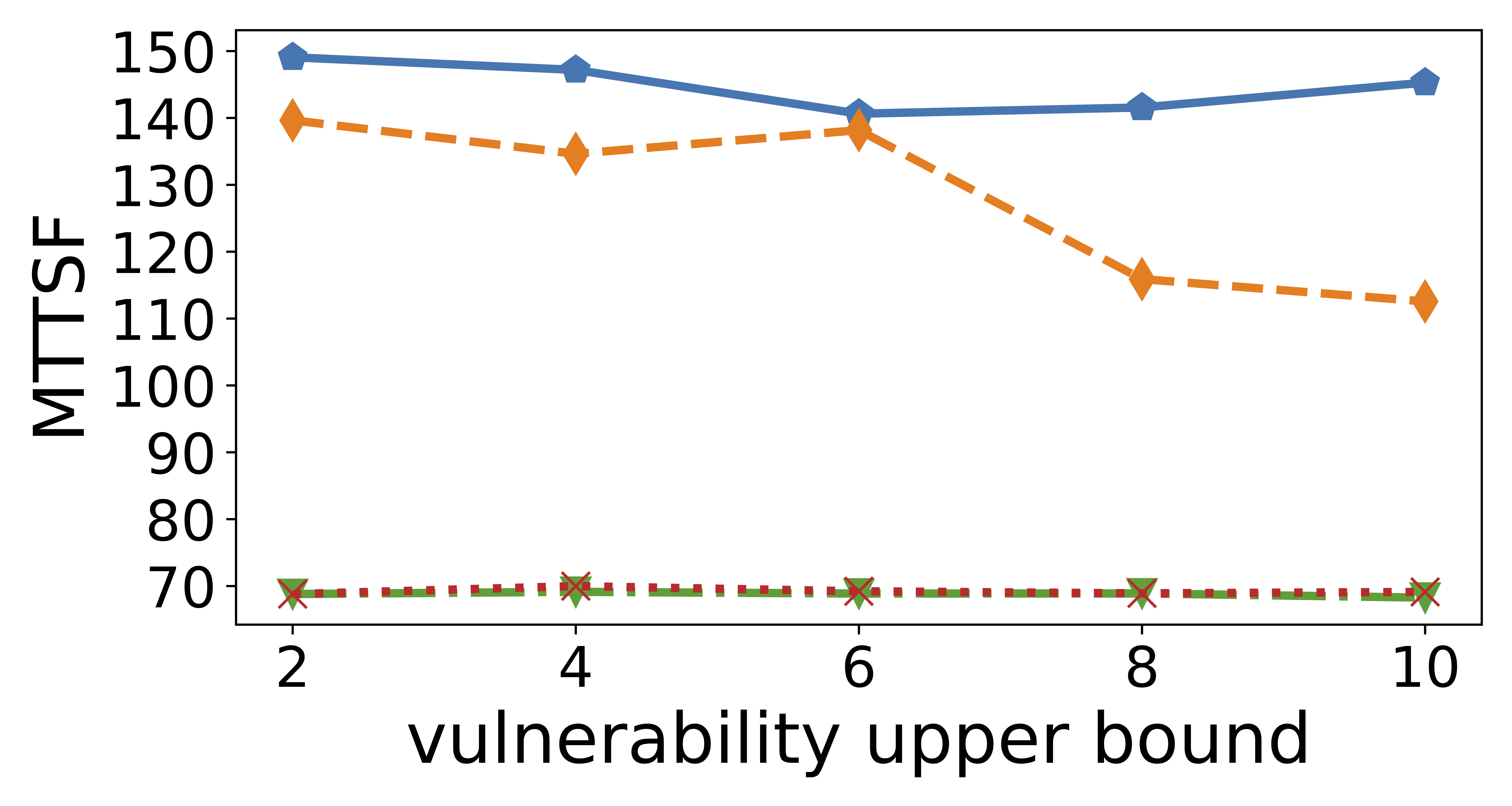}\label{fig: mttsf}}
\hfil
\subfloat[TPR of an NIDS]{\includegraphics[width=.24\textwidth, height=0.2\textwidth]{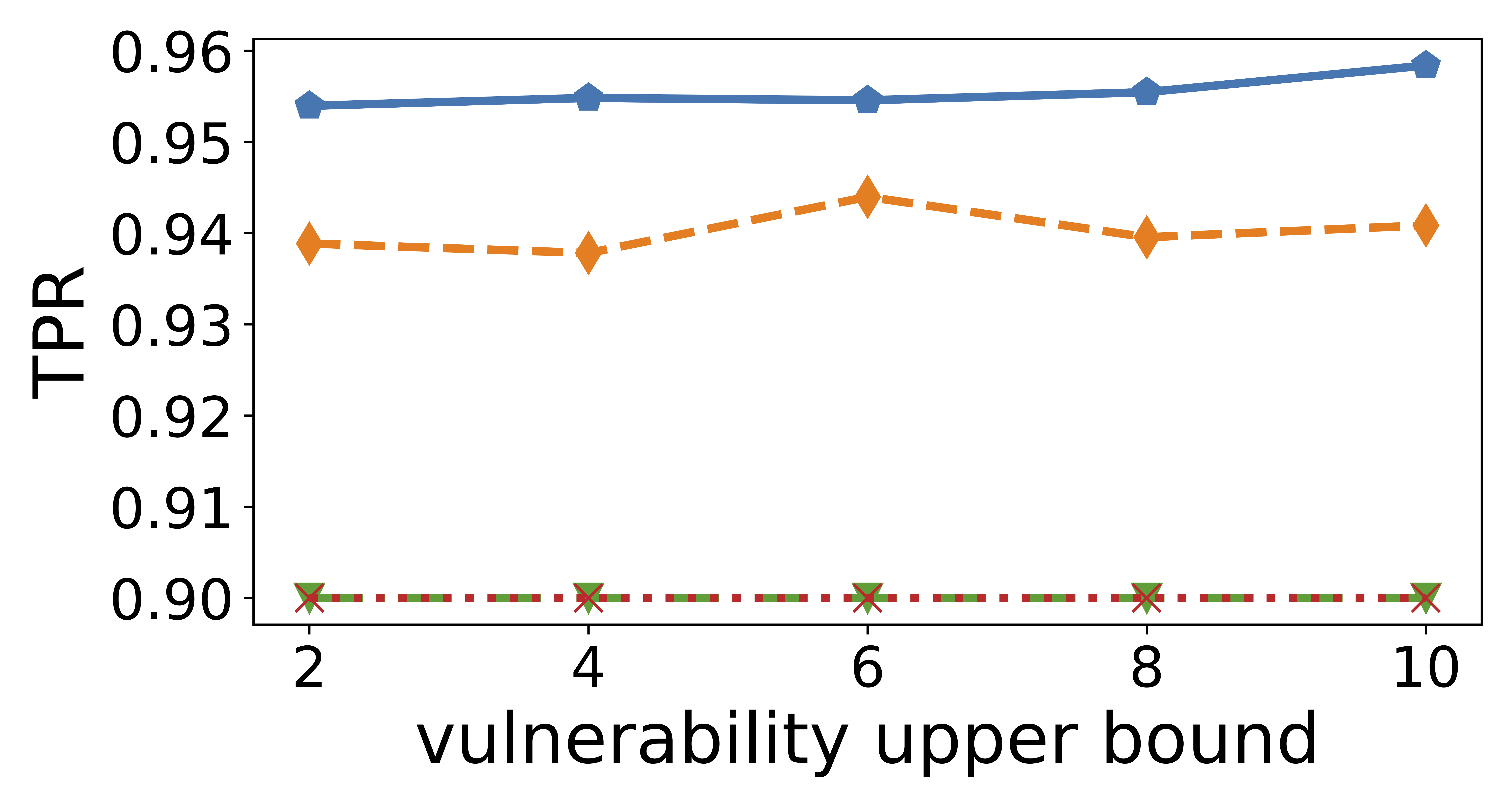}\label{fig: TPR}} 
\caption{System lifetime (i.e., MTTSF) and true positive rate (TPR) of an NIDS under varying the level of system vulnerability. }
\label{fig:TPR-mttsf}
\vspace{-5mm}
\end{figure}

\section{Results \& Analyses} \label{sec:results-analysis}


In Fig.~\ref{fig: attack-uncertainty}, the attacker's perceived uncertainty is plotted for the four defense schemes. When imperfect information (IPI) is considered, the attacker has fairly high uncertainty at the beginning regardless of whether defensive deception (DD) is used or not (i.e., DD-IPI starting from over 0.7 and No-DD-IPI starting from over 0.55). The reason is that using DD strategies can provide a high chance to increase the attacker's uncertainty by misleading the attacker. On the other hand, under perfect information (PI), the attacker's uncertainty is zero.  However, without DD (i.e., No-DD-PI and No-DD-IPI), the system lifetime (i.e., MTTSF) is short, so the curve with respect to the number of games stops at around 80 rounds. Under DD-IPI, the system lifetime much more prolongs compared to No-DD schemes. However, the attacker's uncertainty under DD-IPI decreases with more rounds of games played, because playing more games will result in more compromised nodes. A new attacker can leverage this situation to get into the system quicker. This makes the inside attacker stay longer in the system, resulting in lowering uncertainty as more rounds of games are played.  Some more fluctuations in later games are due to the small number of runs in simulation show long system lifetimes. Notice that using DD makes the system prolong even if it allows some compromised nodes to reside in the system. This is because the system does not evict detected intrusions immediately after detecting them but does reassess them to reduce false positives while collecting more attack intelligence.  This process can give a chance for the NIDS to improve its detection rate. In addition, the attacker perceives lower uncertainty as more games are played as it can perceive less uncertainty about the system since it has been in the system for a while. This was intentionally allowed by the defender to collect attack intelligence.      

In Figs.~\ref{fig: attack-heu} and~\ref{fig: attack-cost}, under varying the vulnerability of nodes in the network, we plotted AHEU and attack cost for each defense scheme. We didn't observe any noticeable sensitivity with respect to varying the extent of node vulnerability.  This is because all three metrics, uncertainty, AHEU, and attack cost do not depend on network conditions but rather depend on the choices of strategies by the attacker and corresponding impact and cost in HEU. In terms of AHEU in Fig.~\ref{fig: attack-heu}, overall, the attacker performs better under DD-based schemes than No-DD-based schemes. The reason is that under DD-based schemes, attackers can use more strategies by being an insider of the system while performing only monitoring attacks as an outside attacker under No-DD-based schemes. In addition, this leads the attacker naturally to perform better under PI than IPI. This explains why the attacker obtains the highest AHEU under DD-PI while having the lowest AHEU under No-DD-IPI. In terms of attack cost, the attacker used more cost under IPI while using less cost under PI as shown in Fig.~\ref{fig: attack-cost}. Under uncertainty, the attacker cannot choose its optimal, cost-effective strategy. Moreover, the attacker paid higher cost under DD while incurring a lower cost under No-DD. This implies that DD strategies are effective to mislead the attacker to choose less cost-effective strategies by increasing its uncertainty. We also discussed how the attack cost and AHEU with respect to the number of games in Fig.~\ref{fig:cost-heu} (a)-(b) in Appendix B of the supplement document with the detailed explanations of the observed trends.


In Fig.~\ref{fig: def-uncertainty}, the defender's uncertainty is shown with respect to the number of attack-defense hypergames. Overall under IPI, the defender's uncertainty is much lower than the attacker's uncertainty.  This is because the defender can collect more attack intelligence while the attacker can be interrupted by DD strategies which increase the attacker's uncertainty.  In addition, there are more fluctuations under No-DD-IPI  is because attackers are not allowed to stay in a system if they are detected as compromised. This keeps resetting the time the attacker has stayed in the system, which makes the defender observe the same attacker for a longer time.

In Figs.~\ref{fig: def-heu} and~\ref{fig: def-cost}, we further show DHEU and defense cost for varying the vulnerability upper bound of nodes ($U_v$) in the network.  In Fig.~\ref{fig: def-heu}, compared to AHEU (i.e., 2.5 to 6), we can observe much higher DHEU (i.e., 6 to 8.5).  Since HEU is estimated based on the impact and cost of taking a chosen strategy, using DD costs more, leading to lowering DHEU. Besides, under IPI, the defender may not choose its optimal strategy all the time, lowering down DHEU due to less benefit of taking a chosen strategy.  Hence, it is reasonable to observe that the highest DHEU is obtained with No-DD-PI while the lowest DHEU is observed with DD-IPI.  In Fig.~\ref{fig: def-cost}, as expected, the highest defense cost incurs under DD-IPI while the lowest defense cost is observed under No-DD-PI. This also reflects the role of the defense cost in DHEU. DHEU and defense cost with respect to the number of games are also discussed in Fig.~\ref{fig:cost-heu} (c)-(d) in Appendix B of the submitted supplement document. 

In Fig.~\ref{fig: mttsf}, we showed how the four different schemes perform under varying the extent of node vulnerability in terms of MTTSF.  Regardless of whether PI or IPI is considered, DD-based schemes outperformed non-DD-based schemes. Again this is because DD-based schemes allow for the reassessment of detected intrusions, leading to reduction in false positives while improving TPR of the NIDS.  However, DD-IPI outperformed all other schemes in terms of MTTSF. This is because IPI can allow the defender to effectively leverage the nature of DD strategies for misleading the attacker effectively and making it choose non-optimal strategies.  Moreover, we notice that the behavior of DD schemes is sensitive to node vulnerability, showing the reduced MTTSF under high vulnerability because the attacker can better exploit vulnerable nodes and more efficiently compromise them.  Except insensitivity under high vulnerability nodes, the performance trends in TPR of the NIDS are well aligned with those in MTTSF under the four schemes, as shown in Fig~\ref{fig: TPR}.  TPR can be improved under DD-IPI due to the high effectiveness of DD under IPI. 

We also discussed the probability of each strategy taken by an attacker and a defender in Figs.~\ref{fig:attack-stra-prob} and \ref{fig:def-stra-prob} and TPR of the NIDS in Fig.~\ref{fig:TPR-NG} of Appendix B with respect to the number of attack-defense games played under each scheme in the submitted supplement document.

\section{Conclusion \& Future Work} \label{sec:conclusion-future-work}

From this study, we obtained the following {\bf key findings}:
\begin{itemize}[leftmargin=*, noitemsep]
\item An attacker's and defender's perceived uncertainty can be reduced when defensive deception (DD) is used. This is because the attacker perceives more knowledge about the system as it performs attacks as an inside attacker. On the other hand, the defender's uncertainty can be reduced by collecting more attack intelligence by using DD while allowing the attacker to be in the system.
\item Attack cost and defense cost are two critical factors in determining  HEUs (hypergame expected utilities). Therefore, high DHEU (defender's HEU) is not necessarily related to high system performance in MTTSF (mean time to security failure) or TPR (true positive rate) which can also be a key indicator of system security.  Therefore, using DD under imperfect information (IPI) yields the best performance in MTTSF (i.e., the longest system lifetime) while it gives the minimum DHEU among all schemes.

\item DD can effectively increase TPR of the NIDS in the system based on the attack intelligence collected through the DD strategies.
\end{itemize}

This work bring up some important directions for future research by: (1) considering multiple attackers arriving in a system simultaneously in order to consider more realistic scenarios; (2) estimating each player's belief based on machine learning in order to more correctly predict a next move of its opponent; (3) dynamically adjusting a risk threshold, i.e., Eq.~\eqref{eq:detect-risk}, depending on a system's security state; (4) introducing a recovery mechanism to restore a compromised node to a healthy node allowing the recovery delay; (5) developing an intrusion response system that can reassess a detected intrusion in order to minimize false positives while identifying an optimal response strategy to deal with intrusions with high urgency; and (6) considering another intrusion prevention mechanism, such as moving target defense, as one of the defense strategies.

\section*{Acknowledgement} This research was partly sponsored by the Army Research Laboratory and was accomplished under Cooperative Agreement Number W911NF-19-2-0150. In addition, this research is also partly supported by the Army Research Office under Grant Contract Number W91NF-20-2-0140. The views and conclusions contained in this document are those of the authors and should not be interpreted as representing the official policies, either expressed or implied, of the Army Research Laboratory or the U.S. Government. The U.S. Government is authorized to reproduce and distribute reprints for Government purposes notwithstanding any copyright notation herein.

\bibliographystyle{IEEETranSN}
\bibliography{hypergame-ref}

\clearpage
\appendix
\begin{appendices}

\section{Hypergame Theory} \label{sec:hypergame}
In this section, we briefly discuss hypergame theory which is mainly leveraged to propose the hypergame theoretic defensive deception framework that deals with APT attacks in this work. This section was mainly used in Section 4 of the main paper.




Hypergame theory offers two levels of hypergames that can be used to analyze games differently perceived by multiple players~\cite{Fraser84}. We adopt first-level hypergames for simplicity. Although hypergame theory applies to multiple players, we consider a game of two players, an attacker and a defender.

\subsection{First-Level Hypergame} 
Given two players, $p$ and $q$, vectors of their preferences, denoted by $V_p$ and $V_q$, define game $G$ that can be represented by $G = \{ V_p, V_q\}$~\cite{Fraser84}. Note that $V_p$ and $V_q$ are player $p$'s and player $q$'s actual preferences (i.e., ground truth), respectively. If all players exactly know all other players' preferences, all players are playing the same game because their view of the game is the same. However, in reality, that assumption may fail. Player $p$ can perceive player $q$'s preferences differently from what they are, leading to differences between $p$'s view and $q$'s view. A game perceived by player $p$ based on its perceived preferences about $q$'s preferences, $V_{qp}$, and the game perceived by player $q$ based on its perceived preferences about $p$'s preferences, $V_{pq}$, can be given by
\begin{equation} \label{eq:perceived_G}
G_p = \{V_{qp}\}, G_q = \{V_{pq}\} \end{equation}
Hence, the first-level hypergame $H$ perceived by each player is written by
$\mathbf{H}^{1} = \{G_p, G_q\}$.
In a first-level hypergame, analysis is performed at the level of each player's perceived game because each player plays the game based on its belief. Even if the player does not know all outcomes of the game, the outcome can be stable for the player because the player may not unilaterally change its belief. If a game includes an unknown outcome, the unknown outcome is caused by the uncertainty. The stability of an outcome about a game is determined {by each player's reaction to the} action by the opponent. An outcome is \emph{stable} for $p$'s game if the outcome is stable in each of $p$'s perceived preference vectors, i.e., in each $V_{qp}$. The equilibrium of $p$'s game is determined by the outcome that $p$ believes to resolve the conflict~\cite{Fraser84}.

\subsection{Hypergame Normal Form (HNF)}
\citet{Vane06} provides a hypergame normal form (HNF) that can succinctly model hypergames based on players' beliefs and possible strategies of their opponents. HNF is formulated, similar to the normal strategic form in game theory. HNF consists of the following four key aspects: (1) full game; (2) row-mixed strategies (RMSs); (3) column-mixed strategies (CMSs); and (4) belief contexts.

The {\bf full game} is the grid form consisting of row and column strategies, which are associated with the utilities, $ru_{11}, \cdots, ru_{mn}$  and $cu_{11}, \cdots, cu_{mn}$ where $m$ is the number of the row player's strategies and $n$ is the number of the column player's strategies. The full game's grid form $\mathbf{U}$ can be represented by an $m \times n$ matrix with an element $ru_{ij}$, $cu_{ij}$ for $i=1, \cdots, m$ and $j=1, \cdots, n$.
\begin{eqnarray} \label{eq:utilities}
\mathbf{U} =
\left(
\begin{tabular}{ccc}
($ru_{11}$, $cu_{11}$)  & $\cdots$ & ($ru_{1n}$, $cu_{1n}$) \\
$\cdots$ & $\cdots$ & $\cdots$ \\
($ru_{m1}$, $cu_{m1}$) & $\cdots$ & ($ru_{mn}$, $cu_{mn}$)
\end{tabular}
\right),
\end{eqnarray}
where $R_0$ and $C_0$ denote the full-game strategies by the row and column
players, respectively.

\vspace{1mm}
\noindent \fbf{Row-mixed strategies} (RMSs) are $m$ strategies the row player considers based on its belief of the column player's strategies. A player's subgame is defined as a subset of the full game (i.e., a set of all possible strategies by all players) because the player may limit a number of strategies it wants to consider based on its belief. Therefore, depending on a situation, the player can choose a subgame to play. RMSs for the $\kappa$-th subgame a player perceives are given by:
\begin{eqnarray} \label{eq:rms}
\text{RMS}_\kappa = [r_{\kappa 1}, \cdots, r_{\kappa m}], \; \; \textrm{where } \sum_{i=1}^m r_{\kappa i} = 1,
\end{eqnarray}
\normalsize
where each probability that a particular strategy $i$ is chosen is estimated by player $p$'s belief based on learning from past experience. Since a subgame consists of a subset of strategies in a full game, if a particular strategy $i$ is not in the subgame $\kappa$, the probability for the row player to take strategy $i$ at subgame $\kappa$ is zero, i.e., $r_{\kappa i}=0$.

\vspace{1mm}
\noindent \fbf{Column-mixed strategies} (CMSs) are a column player's $n$ strategies, believed by a row player for a $\kappa$-th subgame, which are denoted by:
\begin{eqnarray} \label{eq:cms}
\text{CMS}_\kappa  = [c_{\kappa  1}, \cdots, c_{\kappa n}], \; \; \textrm{where } \sum_{j=1}^n c_{\kappa j} = 1,
\end{eqnarray}
\normalsize
where each probability that particular strategy $j$ is chosen is obtained by player $p$'s observations (or learning) towards $q$'s strategies. Similar to the row-mixed strategies, if strategy $j$ is not in subgame $\kappa$, we set $c_{\kappa j}=0$.

\vspace{1mm}
\noindent \fbf{Belief contexts} are the row player's belief probabilities that each subgame $\kappa$ will be played and are represented by:
\begin{eqnarray} \label{eq:sum_belief_contexts}
P = [P_0, \cdots, P_K], \; \; \textrm{where } \sum_{\kappa=0}^K P_\kappa = 1.
\end{eqnarray}

\normalsize $P_0$ is the probability that the full game is played where the full game considers all possible strategies of a player based on the ground truth view of a situation. If the row player is not sure of what subgame $\kappa$ to played due to perceived uncertainty, the unknown belief probability is treated simply for the probability that a full game is played, denoted by 
$P_0 = 1-\sum_{\kappa=1}^K P_\kappa$.

The row player's belief towards the column player's strategy $j$, denoted by $S_h$, is computed by:
\begin{equation} \label{eq:S_i}
S_j = \sum_{\kappa=0}^K P_\kappa c_{\kappa j}~~\textrm{where } \sum_{j=1}^n S_j = 1.
\end{equation}
\normalsize
The summary of the row player's belief on the column player's $n$ strategies is represented by $C_{\sum} = [S_1, S_2, \cdots, S_n]$. 

\subsection{Hypergame Expected Utility} \label{subsec:heu}
The hypergame expected utility ($\text{HEU}$) can be calculated based on $\text{EU}(\cdot)$, and the uncertainty probability perceived by the row player, denoted by $g$, representing the level of uncertainty about what is guessed about a given game. $g$ affects the degree of the $\text{EU}(\cdot)$ of a given hyperstrategy by the row player. $\text{HEU}$ for the given row player's strategy $rs_i$ with uncertainty $g$ is given by~\cite{Vane00}:
\begin{equation} \label{eq:heu}
\text{HEU} (rs_i, g) =  (1-g) \cdot EU (rs_i, C_{\sum}) + g \cdot EU(rs_i, \text{CMS}_w),
\end{equation}
\normalsize
where $rs_i$ is a given strategy $i$ by the row player. $\text{EU} (rs_i, C_{\sum})$ refers to the row player's expected utility in choosing strategy $i$ when the column player can take a strategy among all available strategies $n$. $\text{EU}(rs_i, \text{CMS}_w)$ indicates the row player's expected utility when choosing strategy $i$ when the column player chooses strategy $w$ which gives the minimum utility to the row player. $\text{EU} (rs_i, C_{\sum})$ and $\text{EU}(rs_i, \text{CMS}_w)$ are computed by:
\begin{gather}
\small 
\text{EU} (rs_i, C_{\sum}) =  \sum_{j=1}^n S_j \cdot u_{ij}, 
\label{eq:eu_C}
\\
\text{EU}(rs_i, \text{CMS}_w) = n \cdot S_{w} \cdot u_{iw} 
\label{eq:eu_CMS}
\end{gather}
where $g=0$ means complete confidence (i.e., complete certainty) in a given strategy while $g=1$ implies that the row player is completely occupied with the fear of being outguessed (i.e., complete uncertainty)~\cite{Vane06}.

The calculation of $\text{EU}(rs_i, \text{CMS}_w)$ is based on a pessimistic perspective in that when a player is uncertain, it estimates utility based on the worst scenario. In our work, we consider a realistic scenario in that the player will simply choose a random strategy among strategies in a given subgame. For the defender, when it is uncertain, it will simply play a full game because it does not know what subgame the attacker plays. Note that the utility values will be normalized using min-max normalization method~\cite{han2011data}.

\section{Additional Experimental Results}

\begin{table}[t]
\centering
\caption{\sc Table of Notations}
\label{tab:notations-table}
\vspace{-3mm}
\scriptsize 
\begin{tabular}{|P{0.8cm}|p{5.5cm}|P{1.2cm}|}
\hline
Symbol  &  \multicolumn{1}{c|}{Meaning} & Default \\
\hline
\hline
$\mathrm{ac}_k$ & Cost of attack strategy $k$ (main paper) & Table~2 \\ 
\hline
$\mathrm{dc}_k$ & Cost of defense strategy $k$ (main paper) & Table~3 \\ 
\hline
$\rho_1$, $\rho_2$ & Thresholds for SF in Eq.~(8) in the main paper  & 1/3, 1/2 \\
\hline
$N_{\text{LH}}$ & Number of low-interaction honeypots deployed but not activated in a network & 50 \\
\hline
$N_{\text{HH}}$ & Number of high-interaction honeypots deployed but not activated in a network & 25 \\
\hline
$N_{\text{WS}}$ & Number of Web servers deployed in a network & 25 \\
\hline
$N_{\text{DB}}$ & Number of databases deployed in a network & 25 \\
\hline
$N_{\text{IoT}}$ & Number of IoT nodes deployed in a network & 450 \\
\hline
$N$ & Total number of nodes & 500 \\
\hline
$nv_i$ & Total number of security vulnerabilities of node $i$, including encryption (5), software (5), and unknown (1)  & 11 \\
\hline
$P^r$ & Probability of two nodes being connected in an Erd{\"o}s-R{\'e}nyi random network & 0.05 \\
\hline
$\mathrm{ad}$ & An attacker's deception detectability & $[0, 0.5]$ \\
\hline
$\lambda$, $\mu$ & A constant for normalization for the attacker's uncertainty and defender's uncertainty, respectively & 0.8, 8 \\
\hline
$P_{fp}$, $P_{fn}$ & Probabilities of false positives and false negatives in the NIDS & 0.01, 0.1 \\
\hline
$\mathrm{Th}_{risk}$ & Risk threshold used in Eq.~(6) in the main paper & 0.2 \\
\hline
$\mathrm{Th}_{c}$ & The threshold used in $AS_8$ (Data exfiltration) & 150 \\
\hline
$\epsilon_1, \epsilon_2$ & Increased or decreased percent of a given vulnerability probability by taking attack strategies (i.e., $AS_5, AS_7$) or defense strategies (i.e., $DS_1-DS_3$) & 0.1, 0.01 \\
\hline
$P_{fake}$ & Probability the attacker obtains a fake key & $1-\mathrm{ad}$ \\
\hline
$c_{NT}$ & Percentage (\%) of edges that are hidden by defense strategy $DS_8$ & 20 \\
\hline
\end{tabular}
\vspace{-4mm}
\end{table}

\subsection{Strategy Cost and Hypergame Expected Utility of the Attacker and Defender}

\begin{figure*}[t]
\centering
\subfloat[Attack cost]{\includegraphics[width=.4\textwidth]{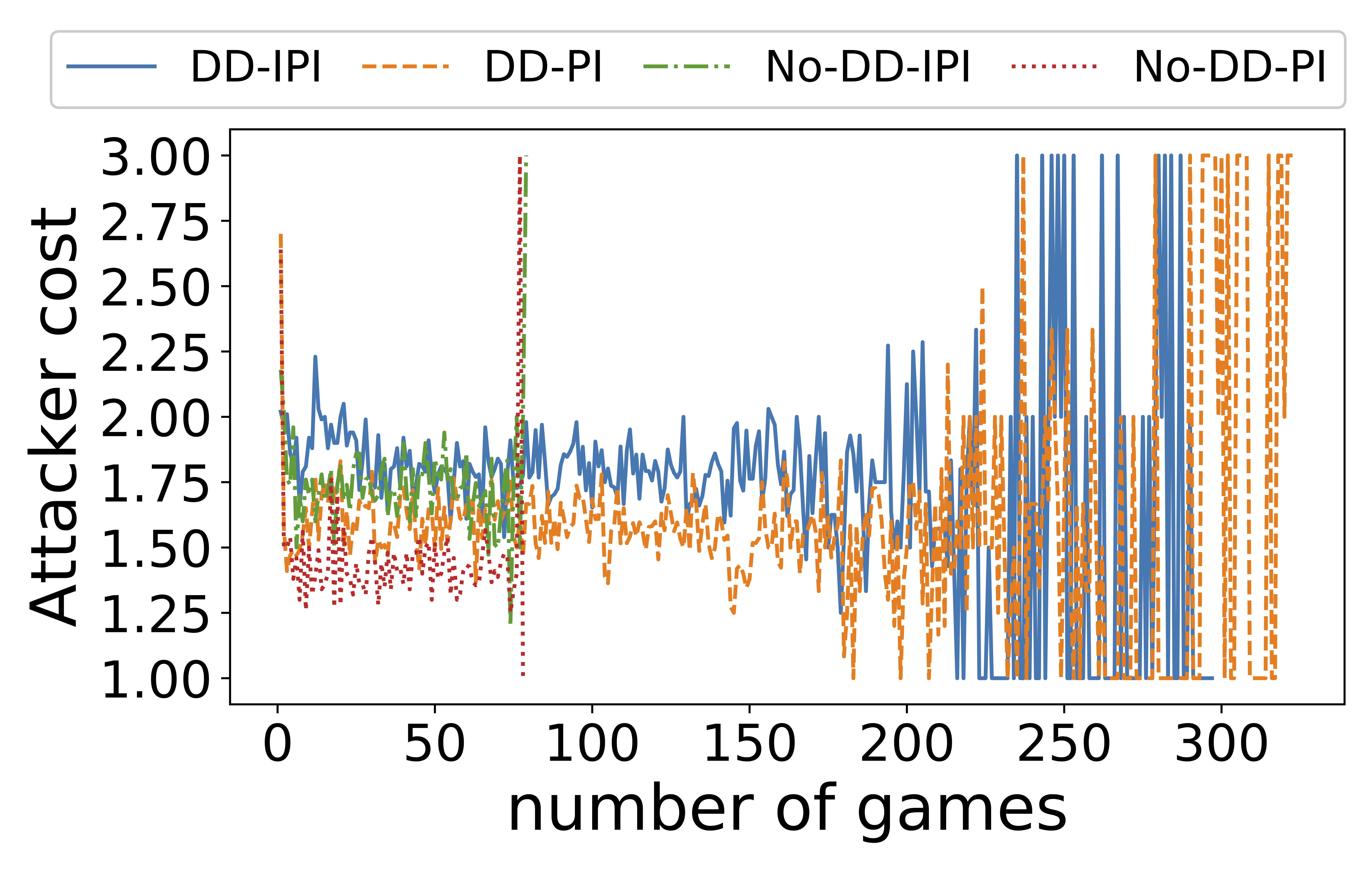}\label{fig: append-attack-cost}}
\hfil
\subfloat[Attacker's HEU]{\includegraphics[width=.4\textwidth]{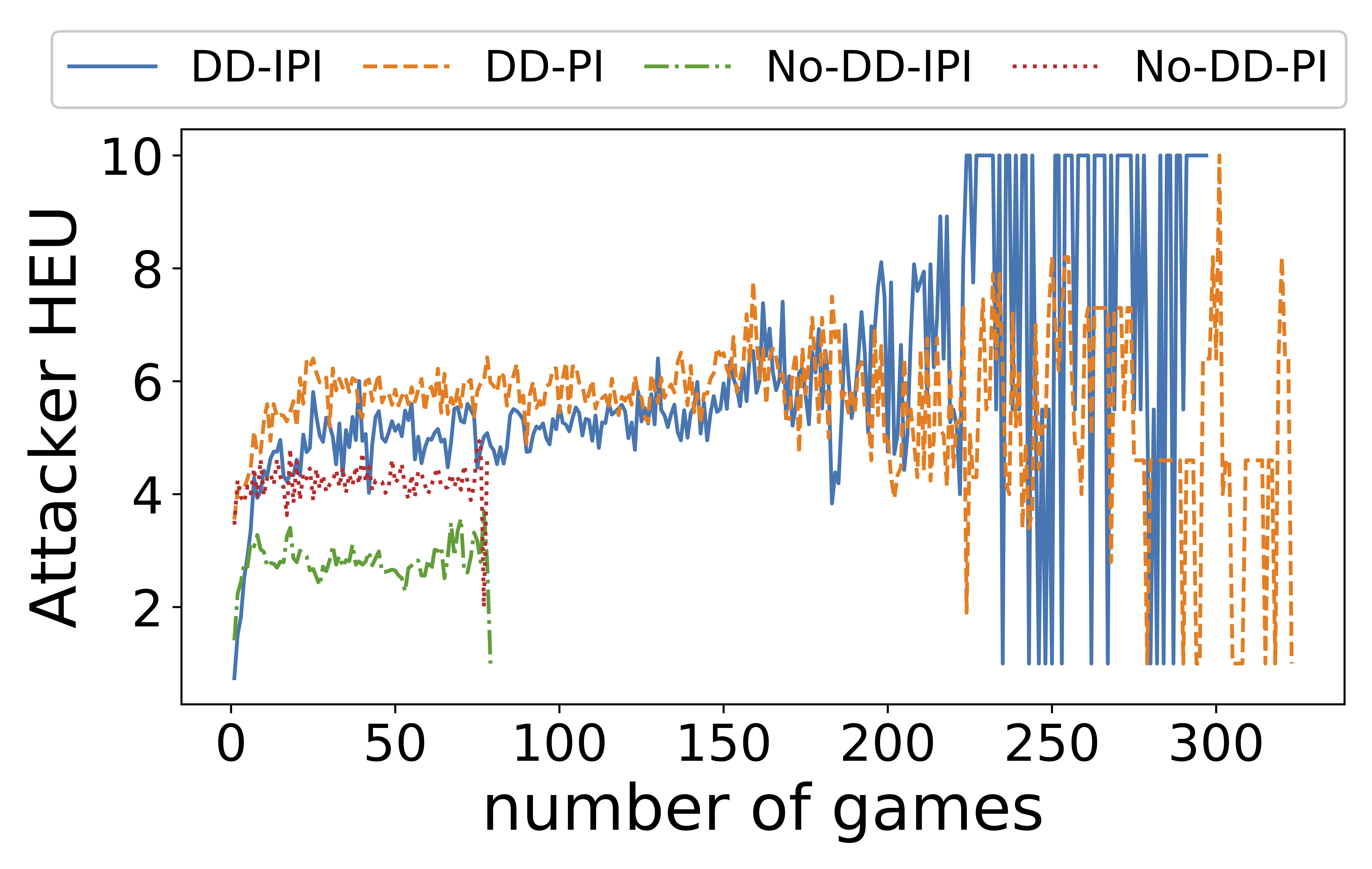}\label{fig: append-attack-heu}} 
\hfil
\subfloat[Defense cost]{\includegraphics[width=.4\textwidth]{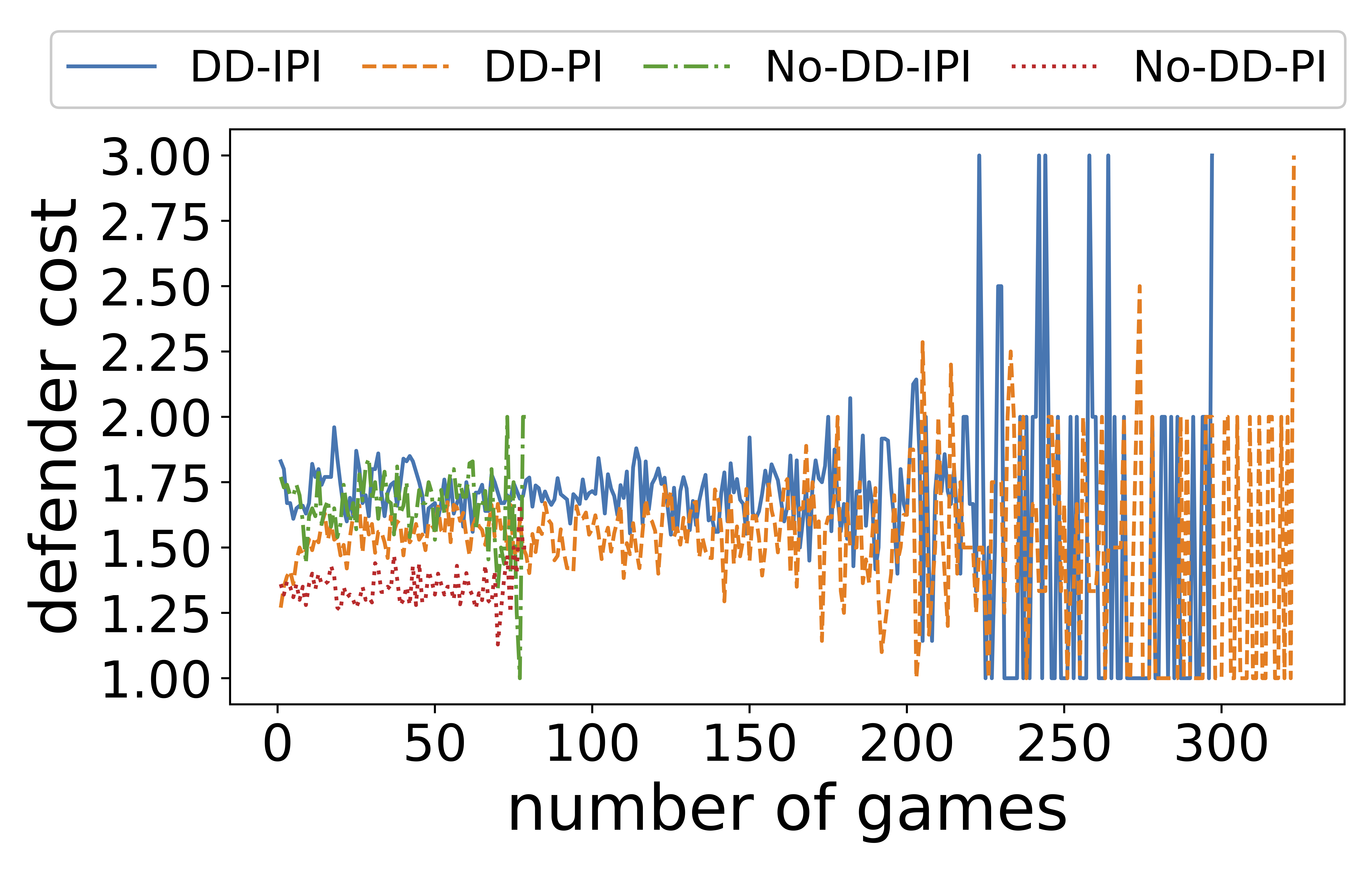}\label{fig: append-def-cost}} 
\hfil    
\subfloat[Defense HEU]{\includegraphics[width=.4\textwidth]{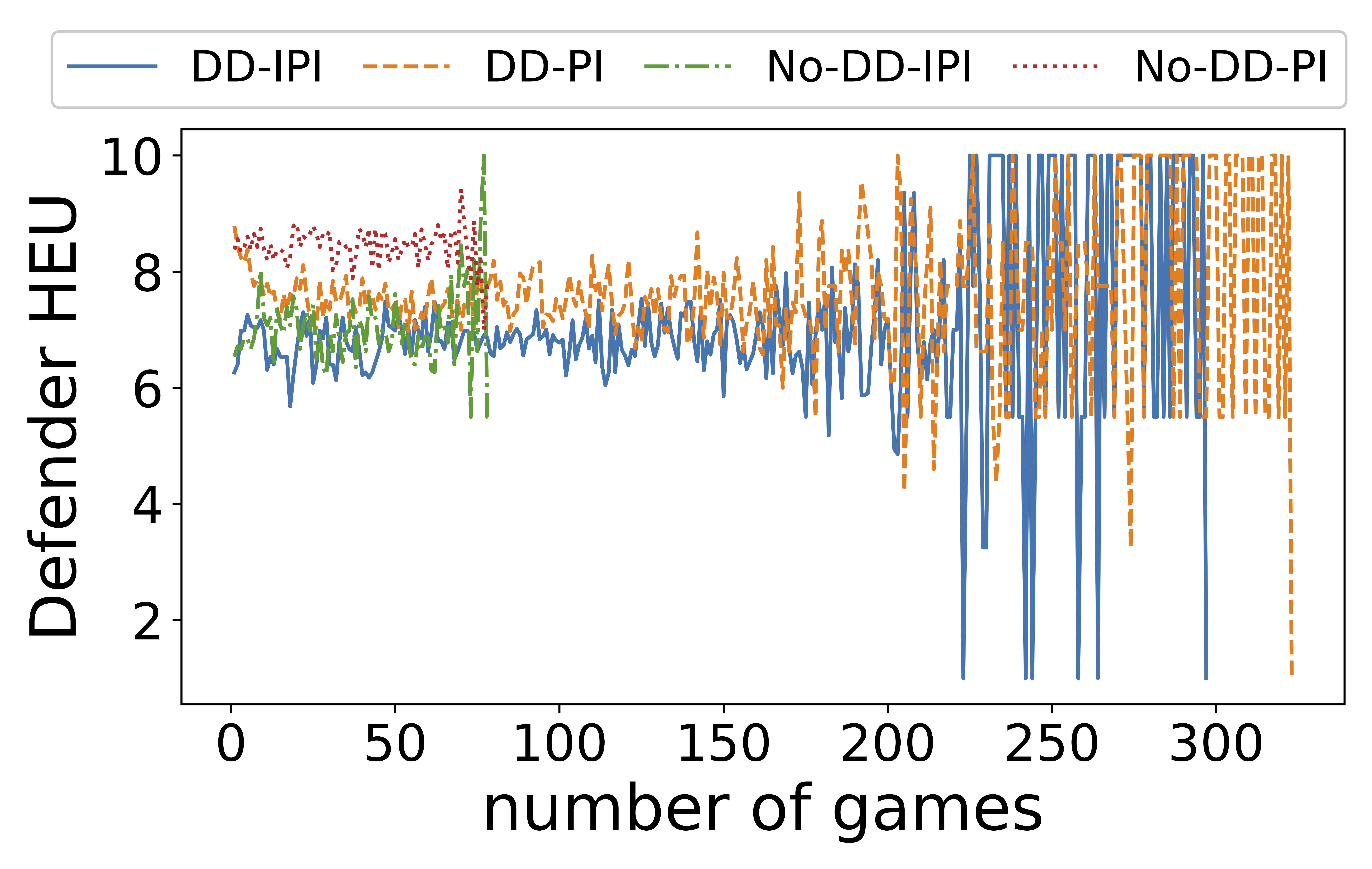}\label{fig: append-def-heu}} 
\hfil    
\caption{Attack cost and defense cost along with the attacker's hypergame expected utility (AHEU) and the defender's HEU (DHEU).  The performance is shown from the second game.}
\label{fig:cost-heu}
\vspace{-3mm}
\end{figure*}

Fig.~\ref{fig:cost-heu} shows the attack cost, defense cost, and the HEUs of the attacker and defender, respectively, with respect to the number of games played between the attacker and defender when the default setting is used based on Table~\ref{tab:notations-table}.  Note that we only showed the number of games from the 2nd game for meaningful analysis.  The main reason of high fluctuations in later games is because only a small number of simulation runs have long system lifetimes, resulting in high variances. We can clearly see the similar trends observed in Fig. 2 of the main paper in that No-DD-based schemes have shorter lifetimes while DD-based schemes show much higher system lifetime, which was measured based on MTTSF in Fig. 4 of this document. Aligned with the trends observed in Fig. 2 of the main paper, in Fig.~\ref{fig: append-attack-cost}, we can observe that when DD-IPI is used, the attacker incurs higher cost than other schemes as it is less likely to choose an optimal strategy, which is cost-effective, due to the confusion or uncertainty introduced by DD-IPI. But under DD-PI, by taking the benefit of perfect information (PI) available, the attacker can take a better action incurring less cost. When No-DD is used, the attacker does not have to use various attack strategies that can be useful as an insider attack because it is less likely to be the inside attacker due to the immediate eviction by the NIDS. In addition, there is no chance for the system to intentionally allow them to be in the system for collecting further attack intelligence. Hence, the attacker can use a limited set of attack strategies that do not incur high cost. 

In Fig.~\ref{fig: append-attack-heu}, we demonstrated the attacker's hypergame expected utility (AHEU) with respect to the number of games played between the attacker and the defender in the default setting.  Under DD-based schemes, when PI is used, higher AHEU is obtained.  On the other hand, imperfect information (IPI) hinders the attacker to choose optimal strategies, leading to less AHEU. This was indirectly exhibited that the proposed DD strategies under uncertainty were effective to confuse the attacker. When No-DD is used, PI helps the attacker to make better attack decisions than IPI. 

Similarly we demonstrated the defense cost in Fig.~\ref{fig: append-def-cost} and the defender's HEU (DHEU) in Fig.~\ref{fig: append-def-heu}.  Under DD-based schemes, in terms of the data availability with respect to the number of games, the system lifetime is observed longer than under No-DD-based schemes with the same reasons explained in the attack cost and AHEU as above. As expected, PI helps the defender to choose better strategies due to no uncertainty perceived than IPI. However, using DD strategies introduces additional cost to achieve better security than using No-DD-based strategies.  Hence, DD-IPI incurs minimum DHEU overall.  However, note that this doesn't mean DD-IPI has less benefit in system security; rather it implies that there is cost to achieve enhanced security.

\subsection{Probabilities of Attack Strategies}

\begin{figure*}[t]
\centering
\subfloat[Under DD-IPI]{\includegraphics[width=.4\textwidth]{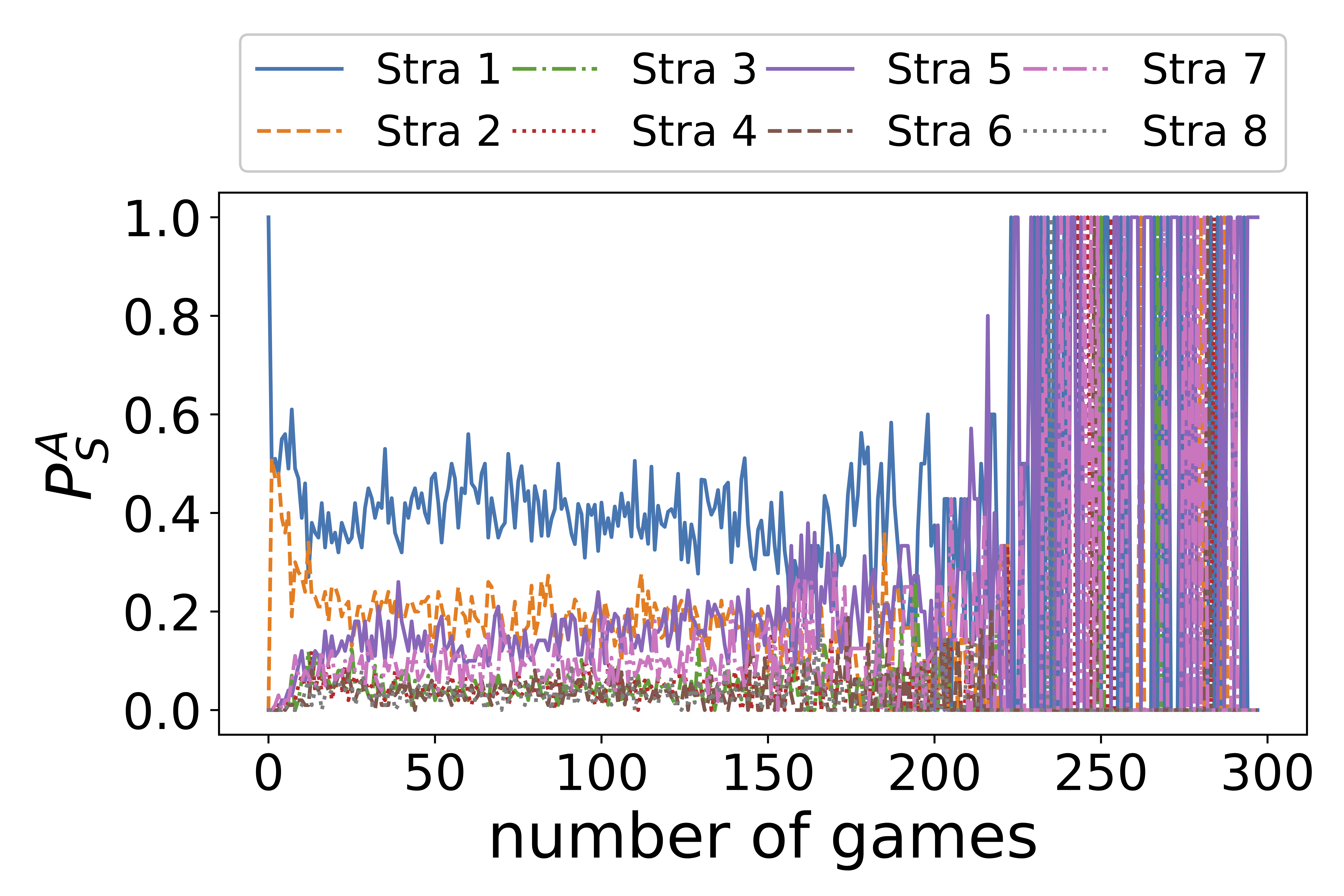}\label{fig: attack-dd-ipi}} 
\hfil
\subfloat[Under DD-PI]{\includegraphics[width=.4\textwidth]{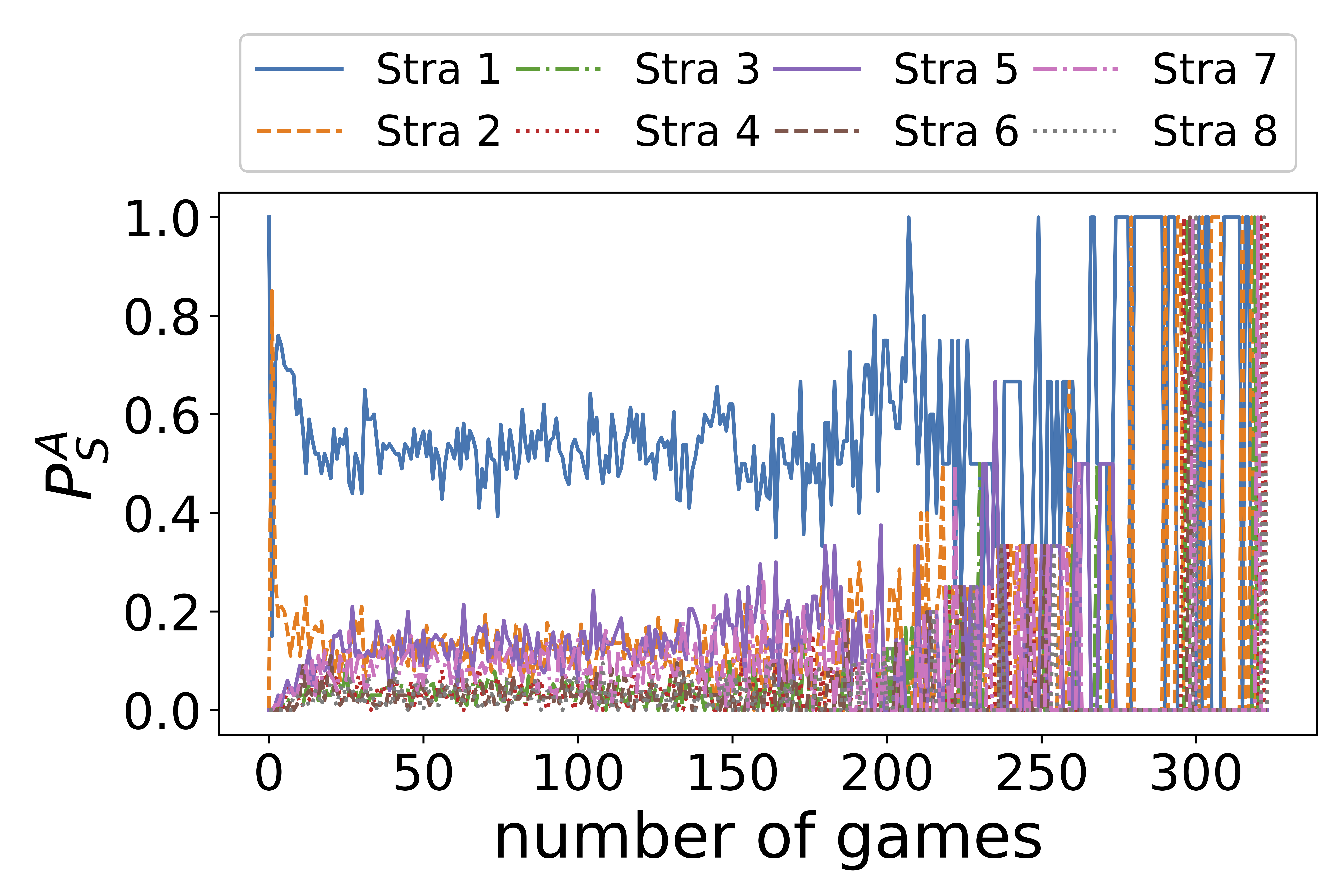}\label{fig: attack-dd-pi}} 
\hfil
\subfloat[Under No-DD-IPI]{\includegraphics[width=.4\textwidth]{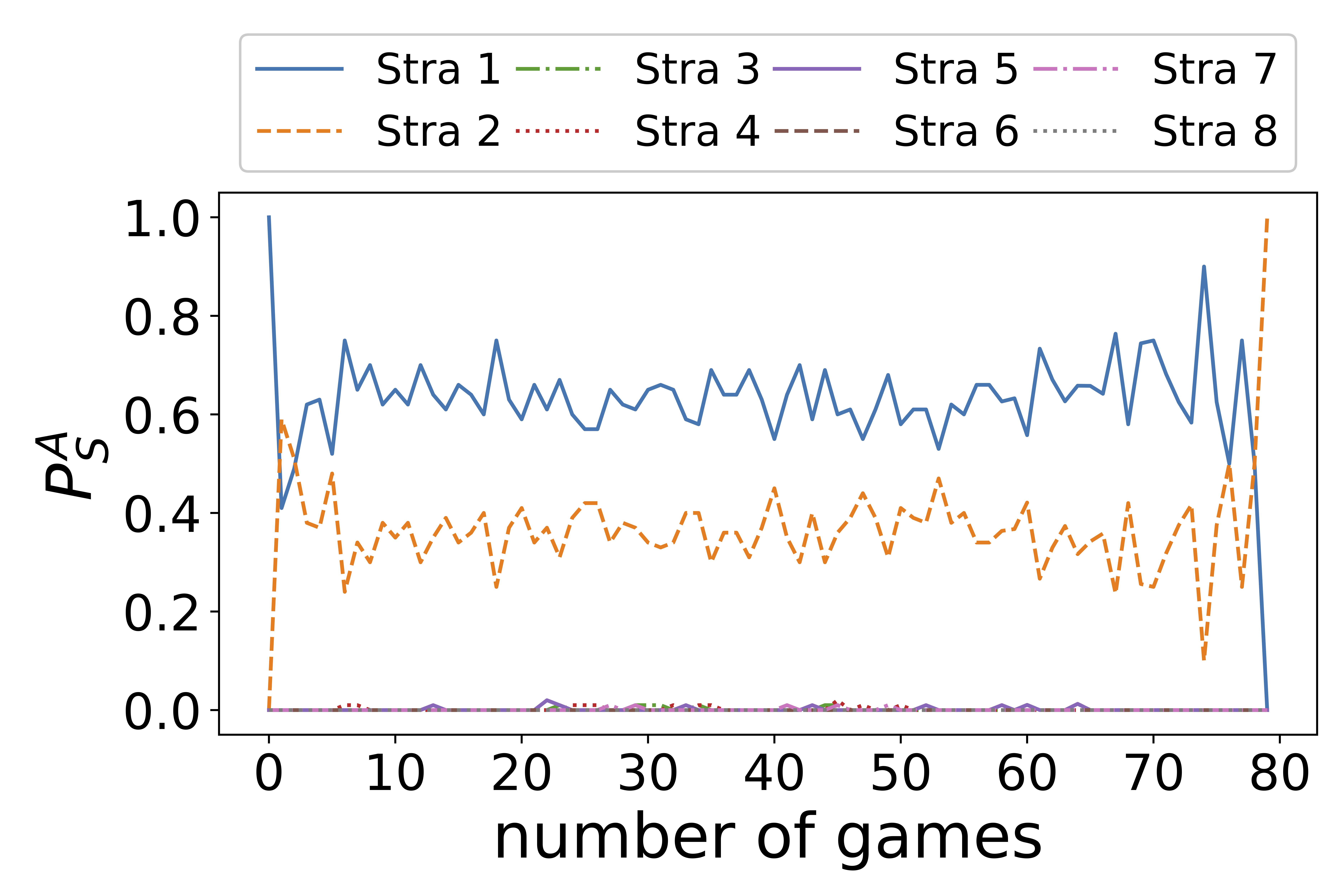}\label{fig: attack-no-dd-ipi}} 
\hfil    
\subfloat[Under No-DD-PI]{\includegraphics[width=.4\textwidth]{./figs/FINAL/No-DD-IPI-att-Strat.png}\label{fig: attack-no-dd-pi}} 
\hfil    
\caption{The probabilities of attack strategies under the four schemes.  The performance is shown from the second game. }
\label{fig:attack-stra-prob}
\vspace{-3mm}
\end{figure*}

Fig.~\ref{fig:attack-stra-prob} shows the probabilities of each attack strategy taken, denoted by $P_S^A$, with respect to the number of games played between the attacker and defender under the four schemes.  Under all the four schemes, attack strategy 1, $AS_1$ (scanning attack), is dominantly taken. This is because every attacker starts from scanning a target system in the stage of reconnaissance (R) in the cyber kill chain (CKC), which is the first step of the advanced persistent threat (APT) attacks.  In addition, due to the presence of the NIDS, when the attacker successfully penetrated into the system, it is highly likely for the attacker to be detected by the NIDS with fairly high detection rate (i.e., $P_{pf}=0.01$ and $P_{pn}=0.1$). In addition, only a compromised node or an attacker, which are not detected by the NIDS or passed the risk threshold ($\mathrm{Th}_{risk}$;  see Eq.~(6) in the main paper), can only remain in the system. Hence, there won't be many insider attackers compared to outsider attackers. Hence, observing the highest probability using scanning attack ($AS_1$) is natural. In addition, as the attacker tries to get into the system by using social engineering attacks ($AS_2$) exploiting both software and encryption vulnerabilities, it is reasonable to observe the attacker taking $AS_2$ as the second most attack strategy.  In Figs.~\ref{fig: attack-dd-ipi} and \ref{fig: attack-dd-pi}, when DD strategies are used, $AS_5$ and $AS_7$ are commonly taken.  This is because these two attack strategies, $AS_5$ and $AS_7$, relatively incur less cost (1 and 2 for attack costs, respectively).  When No-DD strategies are used, it is reasonable to observe that the attacker mainly uses $AS_1$ and $AS_2$ as it does not have many chances to use other strategies due to being detected by the NIDS. When PI is used, the defender also knows more about the attacker due to no uncertainty.  This makes the attacker being evicted quicker and more new attackers are likely to attempt to access the defense system. Hence, the attacker uses social engineering attacks ($AS_2$) more frequently in the later games when PI is used than when IPI is used.

\subsection{Probabilities of Defense Strategies}
\begin{figure*}[t]
\centering
\subfloat[Under DD-IPI]{\includegraphics[width=.4\textwidth]{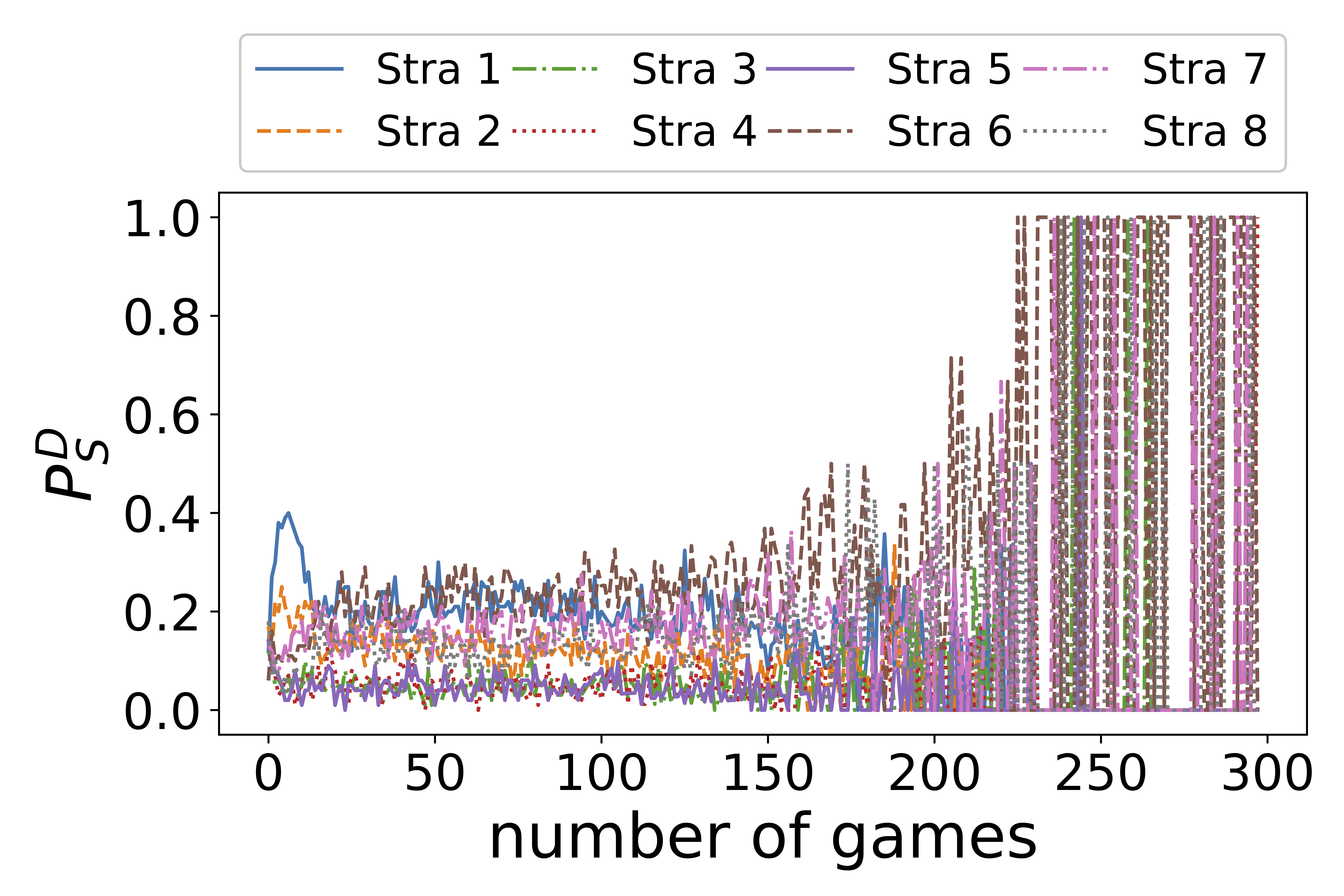}\label{fig: def-dd-ipi}} 
\hfil
\subfloat[Under DD-PI]{\includegraphics[width=.4\textwidth]{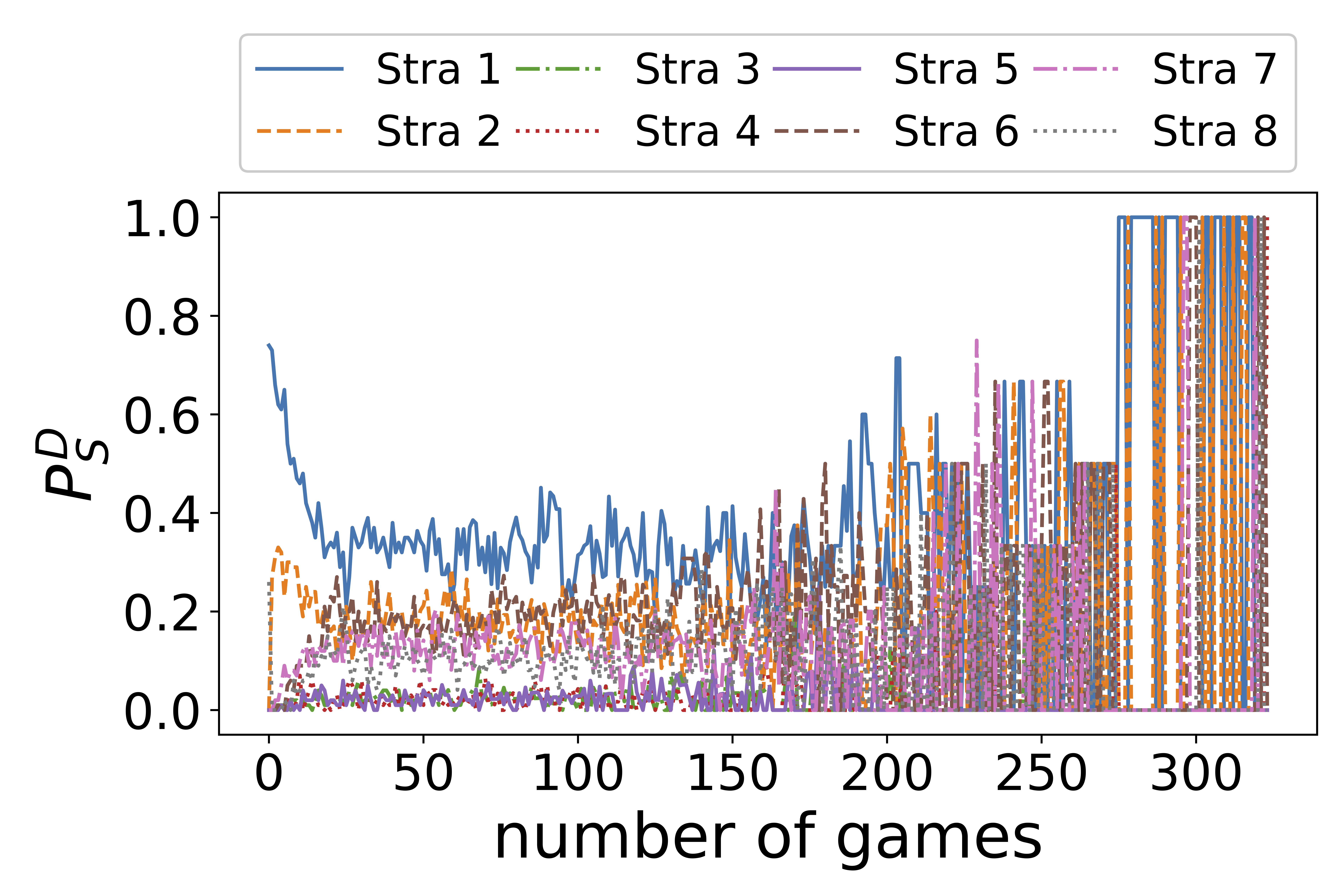}\label{fig: def-dd-pi}} 
\hfil
\subfloat[Under No-DD-IPI]{\includegraphics[width=.4\textwidth]{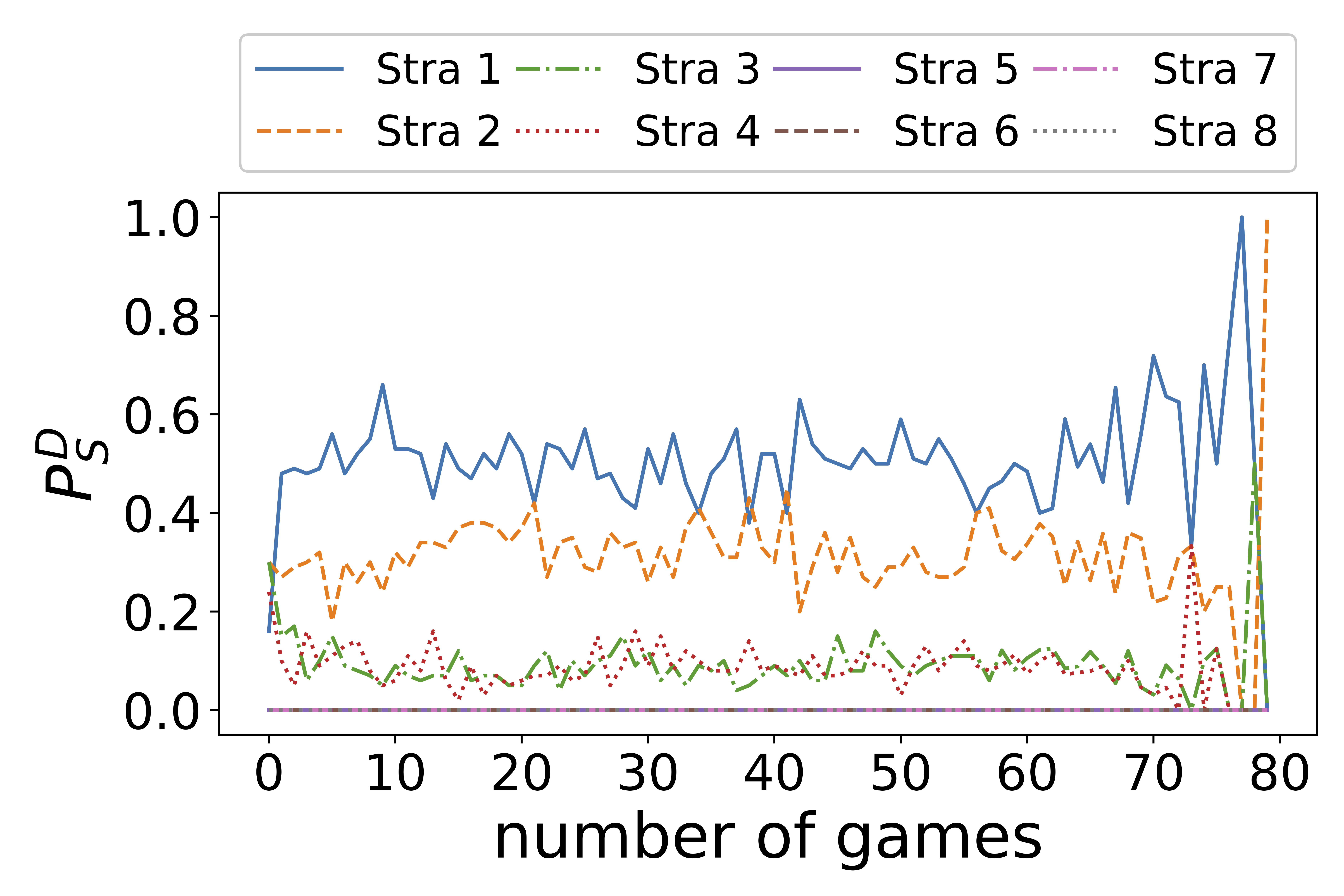}\label{fig: def-no-dd-ipi}}
\hfil    
\subfloat[Under No-DD-PI]{\includegraphics[width=.4\textwidth]{./figs/FINAL/No-DD-IPI-def-Strat.png}\label{fig: def-no-dd-pi}} 
\hfil    
\caption{Probabilities of defense strategies under the four schemes.  The performance is shown from the second game.}
\label{fig:def-stra-prob}
\vspace{-3mm}
\end{figure*}

Fig.~\ref{fig:def-stra-prob} shows the probabilities of the eight defense strategies ($P_S^D$) used under the four schemes with respect to the number of games played between the attacker and defender.  Under all the four schemes, defense strategy 1, $DS_1$ (firewall), is dominantly used as it deals with outside attackers. However, when DD-IPI is used, $DS_2$ (patch management) and $DS_6$ (honey information) were used more commonly compared to other defense strategies.  This is because these two defense strategies cost less, which makes the corresponding DHEU higher, ultimately leading the defender to choose $DS_2$ and $DS_6$ more often than other strategies. When No-DD strategies are used, $DS_1$ and $DS_2$ are mainly used while $DS_3$ and $DS_4$ are marginally used.

\subsection{TPR of the NIDS With Respect To the Number of Games}

\begin{figure}[h]
\centering
\includegraphics[width=.4\textwidth]{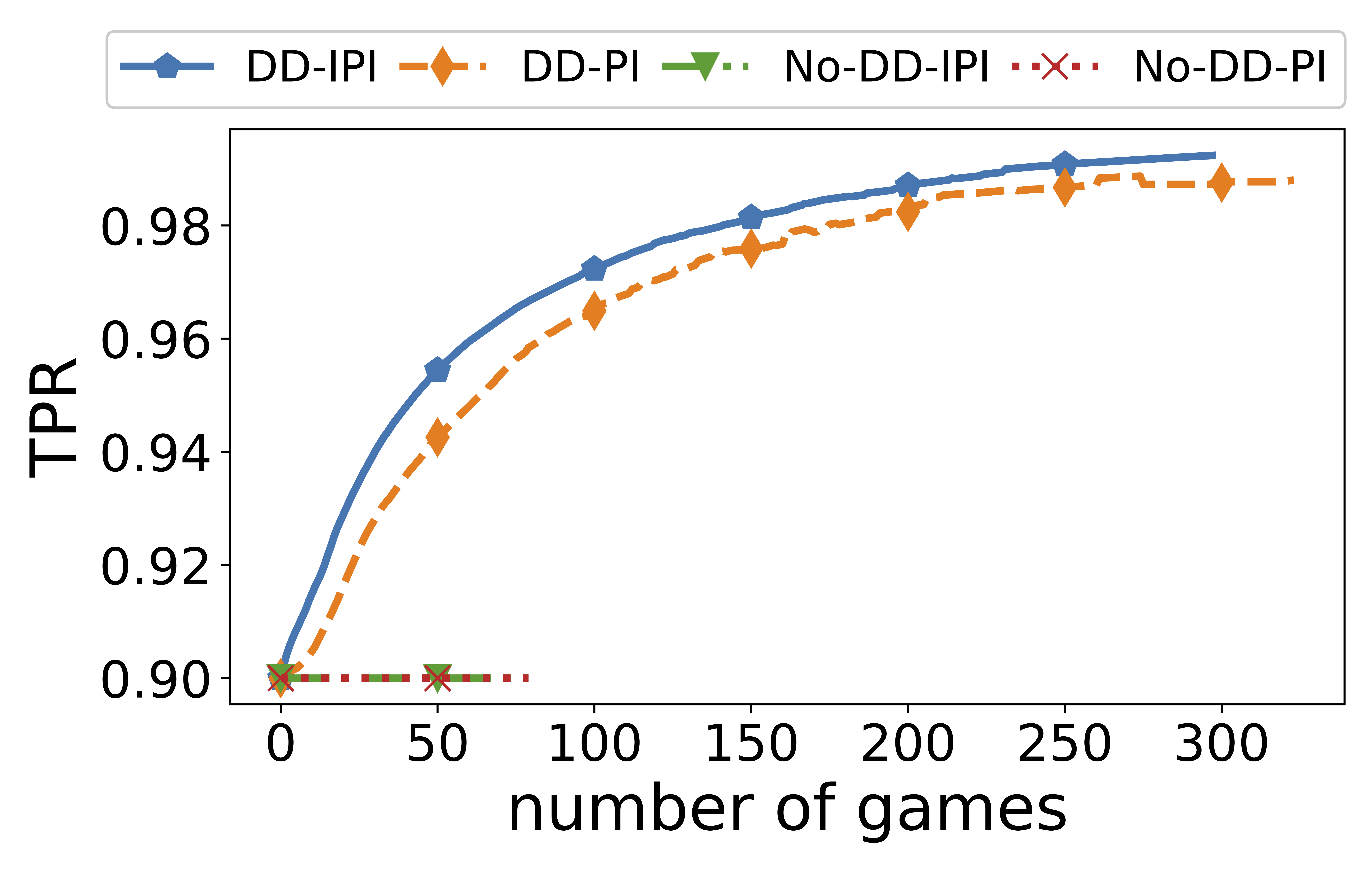}
\caption{True positive rate (TPR) of the NIDS.}
\label{fig:TPR-NG}
\end{figure}

Fig.~\ref{fig:TPR-NG} shows the TPR of the NIDS with respect to the number of games played between the attacker and defender. As expected, since DD-based schemes allow the defender to learn additional attack intelligence which is considered in the NIDS, this naturally leads to the improvement of the TPR in the NIDS.

\section{Representations of Modeled Attack and Defense Strategies, HEUs, and NIDS using the Mind Map}

\begin{figure*}
\centering
\includegraphics[width=.8\textwidth]{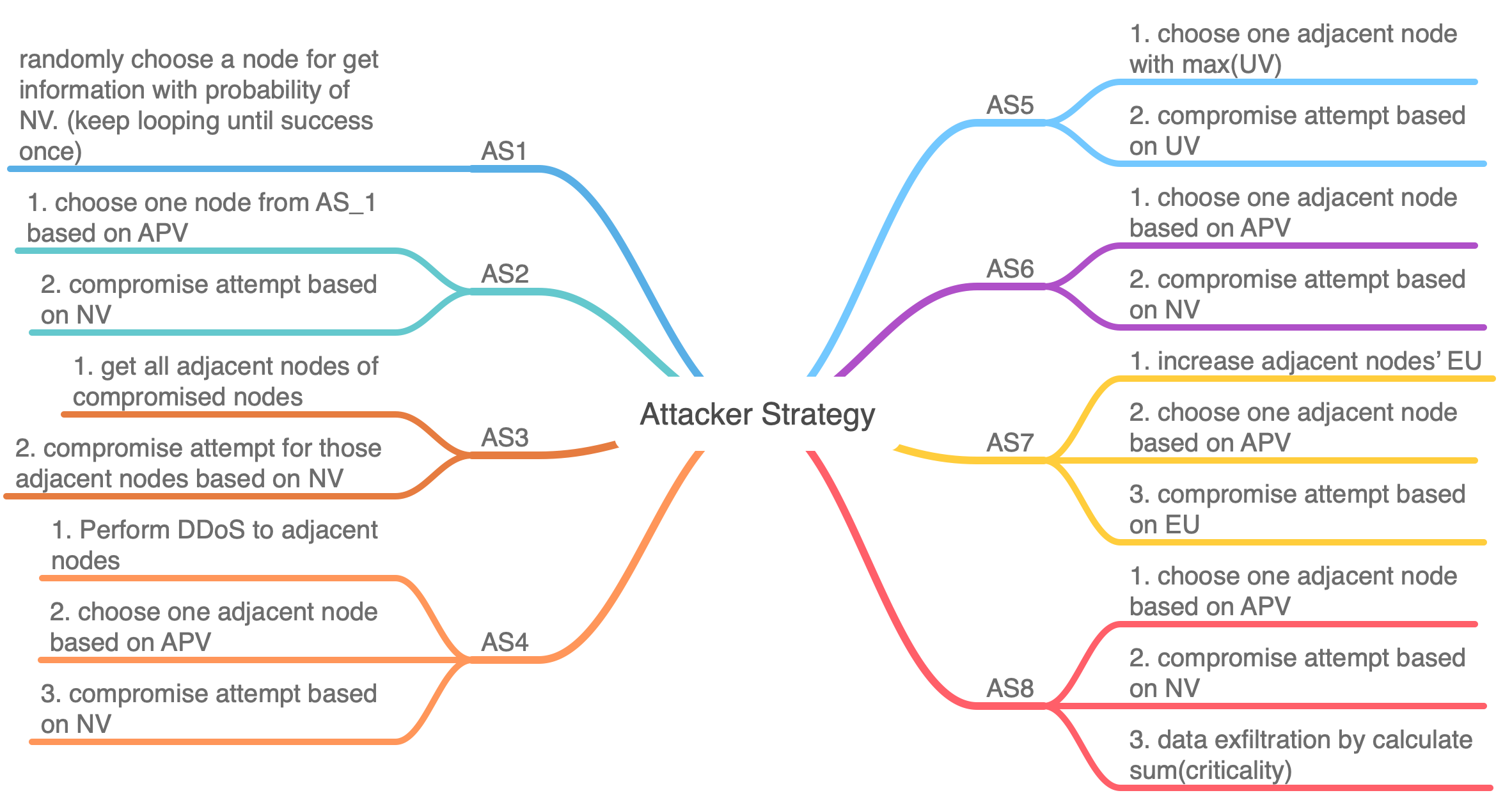}
\caption{Modeling Attack Strategies.}
\label{fig:R-AS}
\end{figure*}

\begin{figure*}
\centering
\includegraphics[width=.8\textwidth]{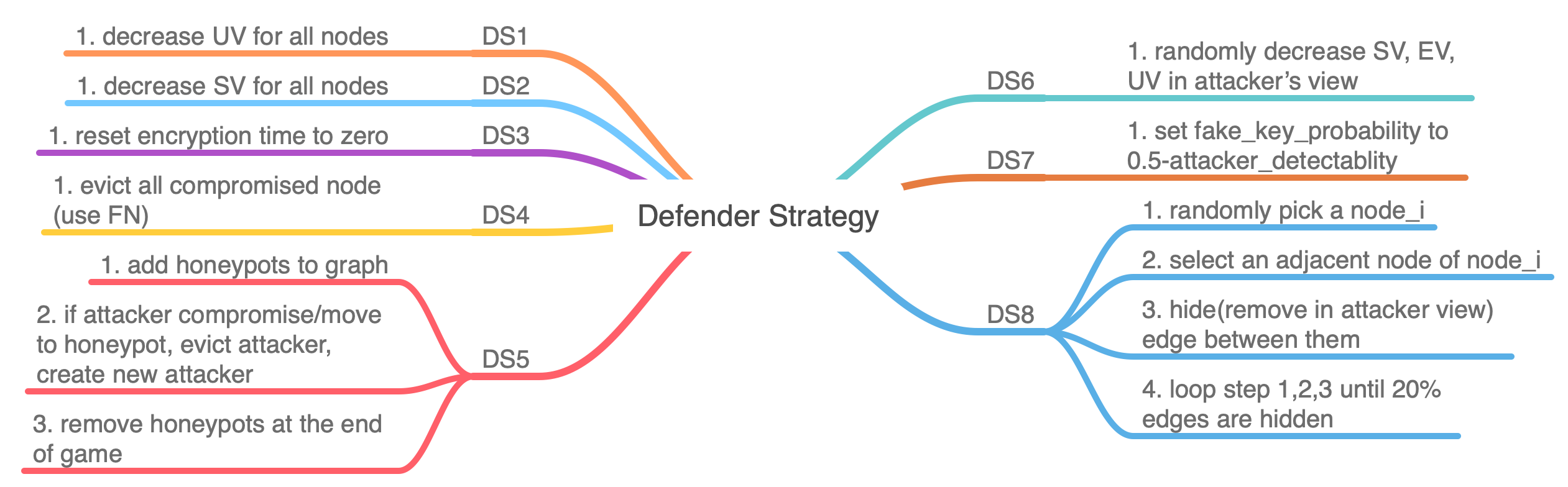}
\caption{Modeling Defense Strategies.}
\label{fig:R-DS}
\end{figure*}

\begin{figure*}
\centering
\includegraphics[width=.8\textwidth]{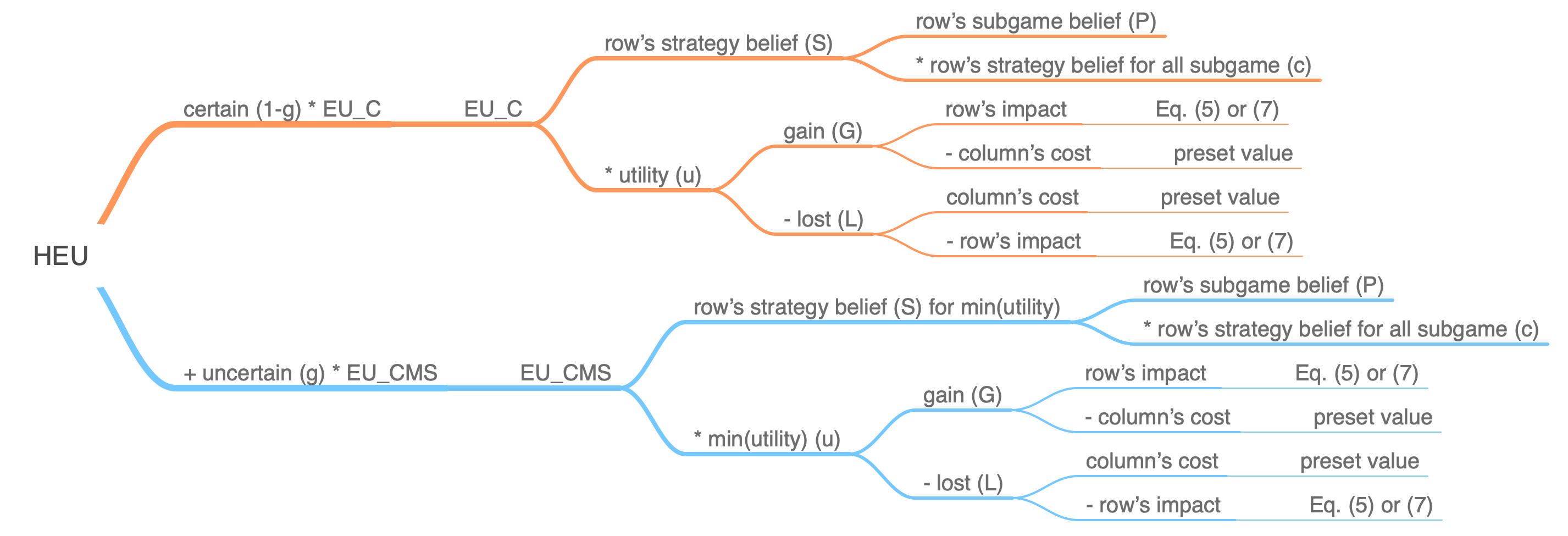}
\caption{Modeling HEU.}
\label{fig:R-HEU}
\end{figure*}

\begin{figure*}
\centering
\includegraphics[width=.8\textwidth]{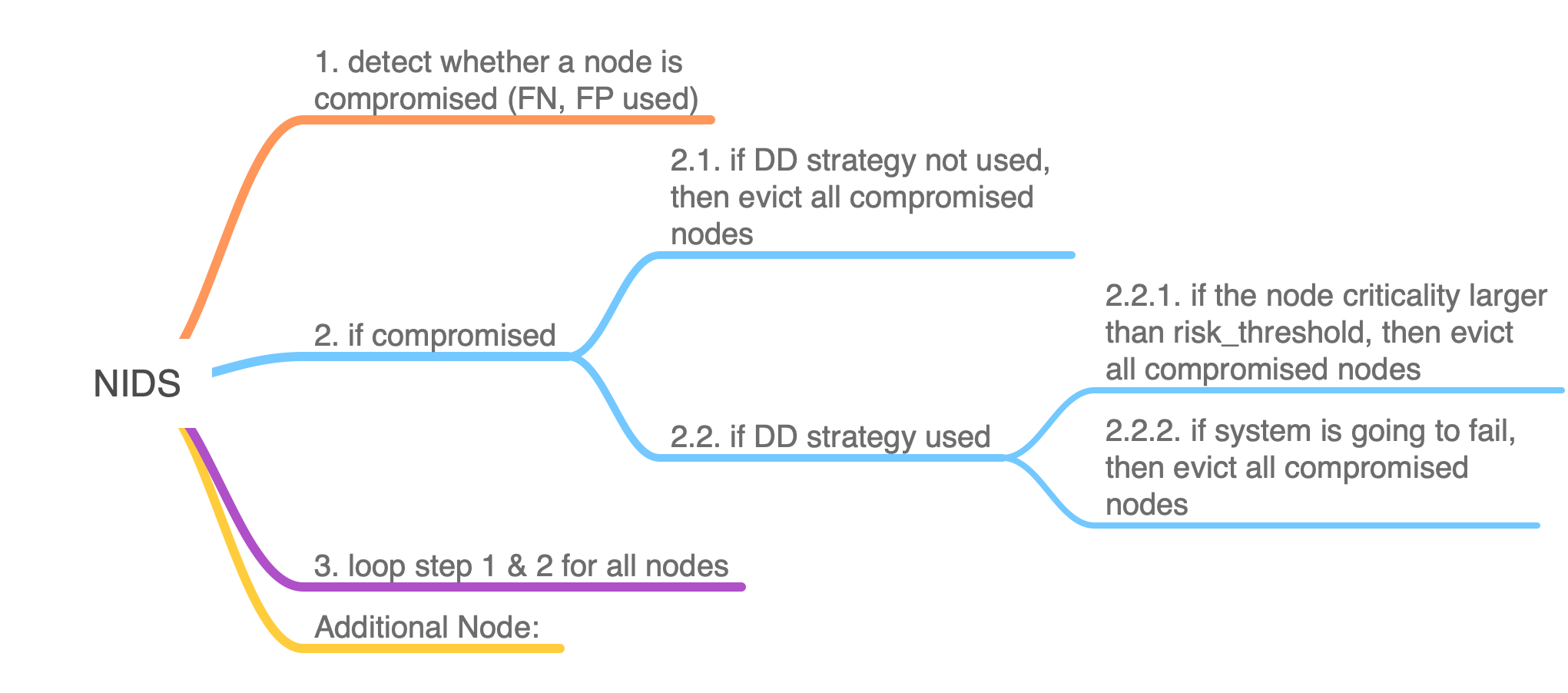}
\caption{Modeling NIDS.}
\label{fig:R-NIDS}
\end{figure*}

For Figs.~\ref{fig:R-AS}-\ref{fig:R-NIDS}, we demonstrated the Mind Maps on how attack and defense strategies are designed, how AHEU and DHEU are estimated, and how the NIDS operates in the given system. These Mind Maps are based on Eqs.~\eqref{eq:cms}-\eqref{eq:eu_CMS} in this supplement document and Eqs.~(9) and (10) in the main paper. For Fig.~\ref{fig:R-NIDS}, we described the workflow on how the NIDS operates in the considered system where each number refers to an execution step. We demonstrated these Mind Maps in order for readers to clearly follow the algorithms used in this work for easy reproducibility.  For the demonstrated Mind Maps, we used the Mind Maps tool called the MindNode~\cite{mindmap}.  

\end{appendices}

\end{document}